\documentclass[a4,11pt]{article}
\usepackage{color}
\usepackage{bm}
\usepackage{slashed}
\usepackage{xcolor} 
\usepackage{graphicx}
\usepackage{amsmath}
\usepackage{amssymb}
\usepackage{amsfonts}
\usepackage{amsthm}
\usepackage{array}
\usepackage[normalem]{ulem} 
\usepackage[sort]{cite}

\usepackage[
      colorlinks=true,
      linkcolor=blue,
      urlcolor=blue,
      filecolor=blue,
      citecolor=red,
      pdfstartview=FitV,
      pdftitle={},
      pdfauthor={},
      pdfsubject={},
      pdfkeywords={},
      pdfpagemode=None,
      bookmarksopen=true
]{hyperref}

\usepackage{epsfig}

\usepackage{hyperref}
\def\aux{\unph A}

\def\beq{\begin{eqnarray}}
\def\eeq{\end{eqnarray}}

\def\b{\beta}

\def\k{\kappa}

\def\D{\Delta}

\def\be{\begin{equation}}
\def\ee{\end{equation}}
\def\bea{\begin{eqnarray}}
\def\eea{\end{eqnarray}}
\def\tilde{\widetilde}

\def\be{\begin{equation}}
\def\ee{\end{equation}}
\def\bea{\begin{eqnarray}}
\def\eea{\end{eqnarray}}

\def\nn{\nonumber}

\newcommand\tf{{\scriptscriptstyle\mathrm{tf}}}

\newcounter{mnotecount}[section]

\renewcommand{\themnotecount}{\thesection.\arabic{mnotecount}}

\newcommand{\mnotex}[1]
{\protect{\stepcounter{mnotecount}}$^{\mbox{\footnotesize
$
\bullet$\themnotecount}}$ \marginpar{
\raggedright\tiny\em
$\!\!\!\!\!\!\,\bullet$\themnotecount: #1} }

\numberwithin{equation}{section}

\newcommand{\phy}[1]{#1}
\newcommand{\unph}[1]{\widetilde{#1}}

\def\Om{\Omega}

\def\dom{\mathcal{D}}

\def\gen{{\unph\xi}}
\def\tgen{\zeta}
\def\spacemet{\gamma}

\usepackage{mathrsfs}
\def\scri{\mathscr{I}}
\def\A{\a (2 \el^{-1} - \Om a)}

\def\man{M}

\def\grOm{\unph{\nu}}
\def\lap{\unph{N}}
\def\unit{\unph{n}}

\def\a{{\tilde a}}

\def\K{\unph{K}}
\def\k{\unph{k}}
\def\ug{\unph{g}}
\def\uspcmet{\unph{\spacemet}}
\def\elec{\unph{E}}
\def\relec{\unph{e}}

\def\el{\ell}

\def\D{D}

\def\X{{\tilde \chi}}

\def\b{{\tilde b}}

\def\phyunit{n}
\def\phylap{N} 
\def\phyk{k}
\def\phyK{K}

\def\tgen{\zeta}
\def\ngen{\zeta_{\perp}}

\newcommand{\const}{\text{const}}

\begin{document}


\title{ \bf{Conformal Einstein equation and symplectic flux with a positive cosmological constant }}
 \vspace{0.8cm} 
\author{  
\bf{Sk Jahanur Hoque}\footnote{jahanur.hoque@utf.mff.cuni.cz (corresponding author)}$~^{1,2,3}$, Pavel Krtou\v{s}\footnote{pavel.krtous@utf.mff.cuni.cz}$~^{1}$, Carlos Peón-Nieto\footnote{carlos.peon@upm.es}$~^{1,4}$
\vspace{0.6cm} \\
$^{1}${\small{\textit{Institute of Theoretical Physics,
Faculty of Mathematics and Physics, Charles University,}}}
\\ 
{\small{\textit{V~Hole\v{s}ovi\v{c}k\'ach 2, 180~00 Prague 8, Czech Republic}}}
\vspace{0.3cm}
\\
$^{2}$ {\small{\textit{Universit\'{e} Libre de Bruxelles, International Solvay Institutes,}}}\\
{\small{\textit{CP 231, B-1050 Brussels, Belgium}}}
\\
$^{3}$ {\small{\textit{Birla Institute of Technology and Science, Pilani, Hyderabad Campus,}}}\\
{\small{\textit{Jawaharnagar, Hyderabad 500 078, India}}}
\vspace{0.3cm}
\\
$^{4}$ {\small{\textit{ETSI Sistemas Informáticos, Universidad Polit\'ecnica de Madrid,}}}\\
{\small{\textit{C. de Alan Turing, s/n, 28031 Madrid, Spain}}}
}
\date{}
\maketitle 
\begin{abstract}
We analyze the conformal Einstein equation with a positive cosmological constant to extract fall-off conditions of the gravitational fields. The fall-off conditions are consistent with a finite, non-trivial presymplectic current on the future boundary of de Sitter. Hence our result allows a non-zero gravitational flux across the boundary of the de Sitter. We present an explicit gauge-free computation to show that the Gibbons-Hawking boundary term, counterterm in the action, and fall-off condition of gravitational field in conformal Einstein equation are crucial to reproduce the finite symplectic flux. 
\end{abstract}

\newpage

\tableofcontents


\section{Introduction}

Ever since LIGO's first detection of gravitational waves \cite{LIGOScientific:2016sjg}, a breakthrough that was itself the result of years of growing activity in the field, the field has continued to grow into one of the most dynamic areas of research. At the theoretical level, the formalism describing gravitational radiation has a long tradition in the zero cosmological constant setting. The first complete formalism describing gravitational radiation in the full non-linear theory is the well-known Bondi-Sachs formalism \cite{Bondi:1960jsa,Bondi:1962px,Sachs:1962wk,Sachs:1962zza}. This is based on the existence of a particular coordinate gauge which allows one to define the energy flux of gravitational radiation in terms of an object known as {\it Bondi News}.


Another approach due to Penrose and Newman \cite{Penrose:1965am,Newman:1961qr} is based on a frame formalism  to give a description of radiation at null future (conformal) infinity $\scri^+$. A remarkable result by Geroch \cite{Geroch:1977jn} is the definition of a tensor quantity at $\scri^+$, which is additionally invariant under conformal scalings, and that matches the { Bondi News}. This therefore receives the name of {\it News tensor}. The News tensor depends entirely on the intrinsic quantities defined on $\scri^{+}$. It is worth to remark that the coordinate and conformal invariance of the News makes it an optimal candidate for describing a physical quantity such as the gravitational energy flux.

The results by Wald and Zoupas  in \cite{Wald:1999wa} also provide further mathematical structure to 
understand the News tensor and gravitational radiation. By giving a general definition of conserved 
quantities (for arbitrary field theories) in terms of Hamiltonian or equivalently a symplectic form defined at the boundary manifold, 
they find out that in the case of general relativity with zero cosmological constant with conformal 
infinity $\scri^+$ as a boundary, such {conserved} quantity is actually determined by the News tensor. 
{The non-vanishing nature of the New tensor or equivalently the symplectic structure on the boundary is attributed to the gravitational flux.}
The 
symplectic form depends on the linear fields $ \delta g$ over a certain gravitational field $g$. Thus, in 
order to evaluate it at the conformal boundary, one must perform a conformal rescaling of both the 
background $\unph{g} = \Om^2 g$ and the linear field $ \delta \unph g = \Om^2 \delta g$, then analyze the 
fall-off behaviour of the unphysical linear field $ \delta \unph g$. An essential result in order to 
relate the symplectic structure and the News tensor is the fact that $ \delta \unph g$ vanishes to first 
order at $\scri^+$. This fall-off behaviour was previously proven by Geroch and Xanthopoulos in 
\cite{Geroch:1978ur} and, as mentioned by Wald and Zoupas, it can be seen as a consequence of the rigidity\footnote{Note that this happens only in four spacetime 
dimensions. For a higher dimensional approach see \cite{Hollands:2003ie}.} 
of the geometry at $\scri^+$. {The asymptotic behaviour of the linearized fields given by Geroch and Xanthopoulos have also been explored by A. Ashtekar et. al \cite{Ashtekar:1990gc, Ashtekar:1981bq} to understand the symplectic geometry of radiative modes of general relativity.}

Our aim  in this paper is to address the above problem in the context of the positive cosmological constant~$\Lambda$.   Our approach will  be based on Wald-Zoupas formalism. Namely, by calculating a 
presymplectic potential at $\scri^+$, which in turn determines the symplectic form, we will define a 
conserved quantity that is a good candidate for gravitational energy flux \cite{Wald:1999wa}. The Wald-Zoupas formalism 
has been used in recent literature \cite{Anninos:2010zf,Hoque:2018byx,Kolanowski:2021hwo, Compere:2019bua,Compere:2020lrt,Poole:2021avh,Dobkowski-Rylko:2024jmh} to understand the gravitational radiation in presence of positive 
cosmological constant. Several references \cite{Anninos:2010zf,Poole:2018koa,Compere:2019bua, 
Compere:2020lrt,Poole:2021avh} carry out the analysis in a  particular 
conformal and coordinate gauge, named after Fefferman and 
Graham’s work \cite{AST_1985__S131__95_0, fefferman2008ambientmetric}. This is the conformal gauge in 
which $\nabla \Om$ is a geodesic vector field and Gaussian coordinates $\{\Om,x^i\}$ adapted to $\Om = \const$ foliation. 
In a nutshell, there is an asymptotic formal series expansion in the Fefferman-Graham gauge of the 
unphysical metric  $\unph g_{ab} = \Om^2 g_{ab}$, where $g$ is the physical metric (i.e. solving the Einstein equation). In this gauge, $\unph g$ is written as \cite{Starobinsky:1982mr}
\begin{align} \label{10III25.03}
 \unph g = - \el^2 d \Om^2 + \unph\gamma_{ab} d\tilde{x}^{a}d\tilde{x}^{b}=- \el^2 d \Om^2 + (\unph\gamma^{(0)}_{ab}+\Omega \unph\gamma^{(1)}_{ab}+\Omega^2 \unph\gamma^{(2)}_{ab}+\Omega^3 \unph\gamma^{(3)}_{ab}+\mathcal{O}(\Omega^4)) d\tilde{x}^{a}d\tilde{x}^{b},
\end{align}
where $\el^{-2}={\frac{\Lambda}{3}}$ and $\unph\gamma^{(j)}_{ab}$ are coefficients\footnote{These coefficients can be understood as tensors on the surface $\Om=0$, i.e., at the infinity $\scri^+$, and they are exported into spacetime by the time flow, or, equivalently, using the Gaussian coordinates.} in the $\Om$-expansion of the spatial metric $\unph\gamma$ ``near'' ${\Om = 0}$.
The $\Lambda$-vacuum Einstein equation determines a recursive relation for these coefficients and  in four dimensions\footnote{This can be similarly performed in arbitrary higher dimensions \cite{AST_1985__S131__95_0, fefferman2008ambientmetric}.} one obtains \cite{Mars:2021pri},
\bea
\tilde{\gamma}^{(1)}_{ab}&=&0,\\
\tilde{\gamma}^{(2)}_{ab}&=&\ell^2\big(\tilde{r}^{(0)}_{ab}-\frac{1}{4} \tilde{r}^{(0)} \tilde{\gamma}^{(0)}_{ab}\big)=: \ell^2\tilde{s}^{(0)}_{ab},\\ \label{10III25.01}
\tilde{\gamma}^{(3)}_{ab}&=&-\frac{2 \ell^{2}}{3} \tilde{e}^{(0)}_{ab}.
\eea
Here $\unph \gamma^{(0)}$ is the induced metric at $\scri^+$, $\tilde r^{(0)}_{ab}$ and $\tilde r^{(0)}$ its Ricci tensor and scalar, $\tilde s^{(0)}$ its {three-dimensional} Schouten tensor, and $\relec^{(0)}$ is the rescaled electric Weyl tensor of $\unph g$ {at the boundary}. \emph{All terms in the expansion to infinite order can be written solely in terms of $\unph \gamma^{(0)}$ and $\relec^{(0)}$}. The pair  $(\unph \gamma^{(0)},\,\relec^{(0)})$ can be thus understood as Cauchy data for the spacetime metric specified in the conformal setting at the infinity $\scri^+$.

We also note that in Fefferman-Graham gauge, $\tilde{\gamma}_{ab}^{(3)}$ can also be identified with a holographic stress-energy tensor \cite{deHaro:2000vlm,Skenderis:2002wp,Anninos:2010zf,Kolanowski:2021hwo}
\bea \label{10III25.02}
T_{ab}:=\frac{2}{|\gamma|^{1/2}}\frac{\delta S}{\delta \gamma^{ab}}=\frac{3}{16 \pi G\ell} \tilde{\gamma}_{ab}^{(3)},
\eea
which is trace-free and divergence-free with respect to the boundary metric $\tilde{\gamma}^{(0)}_{ab}$.

In the Fefferman and Graham gauge, the fall-off behaviour of the linear fields is  obtained by perturbing the seed data  $(\unph \gamma^{(0)}, \relec^{(0)})$ of the expansion \eqref{10III25.03}
\begin{equation}\label{foffFG}
\delta \unph g_{ab} \equiv \delta \unph\gamma_{ab} = \delta \unph \gamma^{(0)}_{ab} + \Om^2 \ell^{2}\delta \tilde{s}^{(0)}_{ab} -\frac{2\ell^{2}}{3} \Om^3 \delta \relec^{(0)}_{ab} + O(\Om^4).
\end{equation}
 The above behaviour of the linear fields is general for all conformally extendable metrics. This is a consequence of the strong non-linear stability result that stems from Friedrich's conformal field equations with positive $\Lambda$ \cite{Friedrich:1986qfi} (see also \cite{Friedrich:2014rpa}). By strong non-linear stability we mean that metrics initially close remain close in the long term (wrt suitable Sobolev norms, cf. Appendix \ref{apphyperbolic}), and also that the conformal extendibility property is preserved. This implies that the asymptotic data (i.e. initial data at $\scri^+$) are also close. Interestingly, the asymptotic data coincide with the degrees of freedom of the Fefferman-Graham expansion, namely $(\unph \gamma^{(0)},\relec^{(0)})$. Thus, given a background metric $\unph g$ with data $(\unph \gamma^{(0)},\relec^{(0)})$, a perturbed metric $\unph g + \delta \unph g$ has data $(\unph \gamma^{(0)} + \delta \unph \gamma^{(0)},\relec^{(0)} + \delta \relec^{(0)})$, so  that \eqref{foffFG} follows. 

The result obtained in four dimensions is that the presymplectic structure is determined by the holographic stress-energy tensor at the boundary \cite{Anninos:2010zf, Kolanowski:2021hwo}. From \eqref{10III25.01}, and \eqref{10III25.02}, it follows that the holographic stress-energy tensor is related to rescaled eletric Weyl tensor in Fefferman-Graham gauge, $T_{ab}=-\frac{\ell}{8\pi G} \tilde{e}_{ab}$. Therefore, one expects to rewrite the presymplectic structure in terms of the rescaled electric Weyl tensor. Although this implies a remarkable coordinate independent identification of the symplectic form, there is no guarantee that it is a conformal gauge invariant. This is because kinematical quantities (depending on the lapse and shift of the foliation determined by $\Om$) may appear, which in the Fefferman and Graham gauge vanish. One of the contributions in this paper is to carry out a coordinate and conformal gauge independent definition of the presymplectic potential at $\scri^+$. Our result matches previous calculations of the presymplectic potential in the Fefferman and Graham gauge, and additionally shows its conformal gauge invariant nature. Without imposing any gauge, we have shown that the natural extension of the holographic stress-energy tensor via conformal Einstein equation can be obtained, and one can write the presymplectic structure in terms of electric part of the rescaled Weyl tensor. Our approach is novel, and to the best of our knowledge we are not aware of explicit derivation of symplectic structure of linearized field in de Sitter in terms of rescaled electric Weyl tensor without imposing any 
gauge.

The fall-off behaviour in \eqref{foffFG} is general in the sense that for every conformally extendible physical metric $g$ there exists a Fefferman-Graham gauge such that the linear perturbations of the unphysical metric $\unph g = \Om^2 g$  satisfy \eqref{foffFG}. Although in a strongly gauge dependent way, this solves the problem of analyzing the fall-off behaviour of the linear fields with considerable generality.  There exists, however, the above mentioned interesting approach in the $\Lambda = 0$ case by Geroch and Xanthopoulos \cite{Geroch:1978ur}, which would be interesting to extend to the $\Lambda>0$ case. 
The idea in \cite{Geroch:1978ur} (see Section \ref{seclambdazero} for a review) is to regularize the linearized Einstein equation (with $\Lambda = 0$) for $\delta \unph g$ by first guessing the fall-of behaviour of its components and rescaling them accordingly. Then, by introducing auxiliary fields and defining equations for them, one obtains a regular hyperbolic PDE system in a suitable linear gauge, extending up to $\scri^+$. The fall-off of the linear fields follows then by the well-posedness of the hyperbolic system.

What is remarkable from \cite{Geroch:1978ur} is the relative simplicity of the whole procedure with respect to other approaches such as Friedrich's conformal field equations \cite{Friedrich:1981at, Friedrich:1986qfi}, (see \cite{Kroon:2016ink, Frauendiener:2000mk} for exhaustive review.) In essence, Friedrich framework also consists in supplementing the Einstein equation with additional fields and equations, but the amount of auxiliary equations and fields is  bigger than in \cite{Geroch:1978ur}. Note that, as a counterpart, Friedrich equations work in the full non-linear regime. Linearizing Friedrich's equations is also a possible approach for the problem of determining the decay of the linear fields which, however, will not be considered here.  For a linearized treatment of these equations in the $\Lambda = 0$ case and their application to gravitational radiation theory, we refer the reader to \cite{Feng:2023ppe}.

We wish to investigate the possibility of extending Geroch and Xanthopoulos results \cite{Geroch:1978ur} to $\Lambda >0$ setting. We find that a major difference between $\Lambda = 0$ and $\Lambda >0$ is that in the first case, linear fields can be generally chosen (in a suitable gauge) to always vanish to the first order at $\scri^+$, namely $\delta \unph g = O(\Om)$, while this is not the case for $\Lambda>0$. This appears to be an obstacle to the regularization of the field equation, which seems to prevent the construction of hyperbolic PDE system in the spirit of \cite{Geroch:1978ur}.

Over the last few years, the study of gravitational waves in presence of positive cosmological constant has gained lot 
of attentions. In the context of gravitational wave generation from spatially compact sources, quadrupolar truncated 
linearized solution of gravitational fields around de Sitter background have been obtained in a generalized harmonic 
gauge \cite{Compere:2023ktn, Ashtekar:2014zfa, Ashtekar:2015lla, Ashtekar:2015lxa, Date:2015kma, Date:2016uzr}. For the linearized quadrupolar solution, quadratic flux-balance laws associated with de Sitter isometries have also been obtained \cite{Compere:2024ekl}. By now, 
Bondi-Sachs formalism for gravitational radiation in de Sitter is also well understood \cite{Poole:2018koa,Compere:2019bua,Chrusciel:2021ttc,Chrusciel:2020rlz,Kolanowski:2020wfg,Bonga:2023eml}. Bondi-Sachs formalism also plays a crucial role 
in understanding asymptotic symmetries and memory effect in de Sitter \cite{Compere:2023ktn,Chu:2016qxp,Chu:2016ngc,Tolish:2016ggo,Bieri:2017vni,Jokela:2022rhk}. Recently, there has also 
been work \cite {Geiller:2022vto,McNees:2023tus,McNees:2024iyu,Geiller:2024amx} related to the computation of symplectic potential in partial Bondi 
gauge (the leading order fall-off for gravitational fields are more generic than Bondi gauge) with non-vaninshing 
cosmological constant. 
There are also attempts \cite{Fernandez-Alvarez:2020hsv,Fernandez-Alvarez:2021yog,Fernandez-Alvarez:2021uvz,Fernandez-Alvarez:2023wal,Fernandez-Alvarez:2024bkf} to understand gravitational 
radiation with a  positive cosmological constant in Penrose's conformal completion technique. Our present work contributes in the 
direction of understanding gravitational radiation with a positive cosmological constant in conformal completion 
formalism. In particular, we extract the fall-off behaviour of gravitational fields from conformal Einstein equation and 
show that these fall-off behaviour induces non-zero gravitational flux in the future infinity of de Sitter.

The paper is organized as follows. In Section 
\ref{secregularization} we analyze, in an arbitrary 
conformal gauge, the fall-off behaviour of the linear 
fields with positive $\Lambda$ directly from the 
Einstein equation. In Section \ref{seclambdazero}, 
we study the differences with respect to the $\Lambda = 0$ case. 
Section \ref{Sec:Gauge} revisits the infinitesimal 
diffeomorphisms group from a conformal perspective. 
Then, in Section \ref{secdegfreedom} we study where 
the gauge degrees of freedom are localized within the 
linear fields, which aims to help understanding why 
the gauge \eqref{foffFG} is actually general. We analyze the asymptotic behaviour for background gravitational fields in Section~\ref{seceinstein}. We 
conclude the paper with Section~\ref{secpresymplec}, 
where we apply our previous analysis to compute the 
presymplectic potential in a gauge-independent manner. 
Relevant mathematical tools are given 
in the appendices.

\paragraph{Notation and conventions}

Throughout this paper, we shall consider $4$-dimensional manifold $(\phy M, \phy g)$.
The spacetime metric, as well as their related objects, will be referred to as ``physical'' and always denoted without a tilde. The formalism employed to study asymptotic properties will be that of Penrose's conformal completions. Namely, we consider a smooth positive function $\Om$ nowhere vanishing on $\phy M$, in such a way that the boundary $\scri := \partial \phy M =\{ \Om = 0\}$ must satisfy $d \Om \mid_\scri \neq 0$. We define the conformally rescaled metric
\begin{equation}
 \unph g_{ab} = \Om^2 g_{ab},
\end{equation}
which we require to be smooth on $\unph M := \phy M \cup \partial \phy M$. The manifold $(\unph M ,\unph g)$ is called a conformal extension of $(M,g)$ or simply the ``unphysical manifold". All unphysical objects are denoted with a tilde, and their indices are raised or lowered by $\unph g$.

We will often make use of the foliation defined on $\unph M$ with $\{ \Om = \const\}$ leaves. Therefore, we define the following objects: 
\begin{equation}\label{notation1}
\grOm_a := \unph{\nabla}_a \Om,\quad -\lap^{-2} := \unph{g}^{cd} \grOm_c \grOm_d,\quad \unit_a := \lap \grOm_a,
\end{equation}
where $\lap$ is the lapse function, which we define by the positive root of $\lap^{2}$. 

We now have the following decomposition of the metric $\unph{g}$
\begin{equation}\label{notation2}
\unph{g}_{ab} = -\unit_a \unit_b + \unph{\spacemet}_{ab},
\end{equation}
where $\unph{\spacemet}_{ab}$ is the induced metric at the leaves $\{\Om = \const\}$, with associated Levi-Civita covariant derivative $\unph D$. We introduce also
\begin{equation}\label{notation3}
\K_{ab} := \unph\nabla_a \unit_b,\qquad \k_{ab} := \unph{\spacemet}^{c}{}_{a}\unph{\spacemet}^{d}{}_{b} \K_{cd},
\end{equation}
$\unph k_{ab}$ being the extrinsic curvature of the leaves $\{\Om = \const\}$. Note that typically we assume $\Om>0$ in the physical spacetime $M$. It implies that $\grOm_a$ and $\unit_a$ are inside oriented normal 1-forms, i.e., the past-oriented at the future infinity and the future-oriented at the past infinity, while $\grOm^a$ and $\unit^a$ are outward oriented normal vectors.

We focus on the positive cosmological constant $\Lambda>0$ and we introduce the corresponding cosmological length $\el$,
\begin{equation}
    \el^{-2} = \frac{\Lambda}{3}.
\end{equation}

\section{Fall-off analysis of linearized field in presence of cosmological constant}\label{secregularization}
In this section, we wish to extract a fall-off condition for linearized gravitational fields in presence of positive cosmological constant. For this purpose, we will follow the method of Geroch and Xanthopoulos \cite{Geroch:1978ur}, originally developed for $\Lambda = 0$.
Their approach  (see  Section~\ref{seclambdazero} for a review) regularizes the linearized conformal Einstein equations by first guessing the fall-off behavior of the field components a priori and rescaling them accordingly. Then, by introducing auxiliary fields and imposing suitable gauge conditions, they obtain a well-posed hyperbolic PDE system that extends smoothly to $\scri$. Given the success of this method for $\Lambda = 0$, it is worth exploring its applicability to $\Lambda > 0$.

We will only focus on extracting the fall-off condition of the linearized fields from the conformal linearized Einstein field equation, and 
also explore whether the field variables produce a finite symplectic 
structure at the boundary, $\Omega =0$. The well-posedness of the system of equations for a positive 
cosmological constant is beyond the scope of our paper.

Assume now that the physical metric $g_{ab}$ satisfies 
the Einstein equation for some stress-energy tensor $T_{ab}$,
\be
R_{ab}-\frac{1}{2} R g_{ab}+ \Lambda g_{ab}=8\pi G T_{ab} .
\ee
It is convenient and customary to write down Einstein equation in terms of unphysical Schouten tensor,
\be 
\tilde{S}_{ab}:= \tilde{R}_{ab}-\frac{1 }{6} \tilde{R}\tilde{g}_{ab}.
\ee
Using the conformal transformation between physical and unphysical Ricci tensor we obtain,
\begin{equation}\label{5VII20}
\tilde{S}_{ab}=\big(R_{ab}-\frac{1}{6}R g_{ab}\big) - 2 \Omega^{-1} \tilde{\nabla}_{a} \tilde{\nabla}_{b}\Omega 
         +\Omega^{-2} \  \tilde{g}_{ab} \tilde{g}^{cd} \  \tilde{\nabla}_{c} \Omega \tilde{\nabla}_{d}\Omega.
\end{equation}
Substituting for the curvature terms from the Einstein equation, we get
\begin{equation}\label{17VI24.01}
   \Omega \tilde{S}_{ab} +2 \tilde{\nabla}_{a} \tilde{\nabla}_{b}\Omega 
   - \Omega^{-1} \bigg( \tilde{g}^{cd} \ \tilde{\nabla}_{c} \Omega \tilde{\nabla}_{d} \Omega 
   +\el^{-2} \bigg)\tilde{g} _{ab}
    =8\pi G \Omega^{-1} \bigg(\tilde{T}_{ab}-\frac{1}{3} \tilde{T} \tilde{g}_{ab}\bigg),
\end{equation}
where we define $\unph{T}_{ab}:=\Omega^{2} T_{ab}$. 

Since $\unph{g}_{ab}$ and $\Omega$ are smooth on $\tilde{M}$, the first two terms in the left-hand side of \eqref{17VI24.01} are smooth on $\scri$. 
If the stress-energy tensor on the right-hand side vanishes asymptotically to order one (i.e., $\Om^{-1}\unph{T}_{ab}\equiv\Om T_{ab}$ has a smooth limit on $\scri$), the remaining term on the left-hand side of \eqref{17VI24.01} must be also smooth on $\scri$. We denote it as $\unph f$,
\begin{equation}\label{deff}
  \unph f = \Om^{-1}\big(\unph{g}^{ab} \ \unph{\nabla}_{a} \Om \unph{\nabla}_{b} \Om +\el^{-2} \big) 
    = \Om^{-1}\big(\unph{g}^{ab} \ \unph{\nu}_{a} \unph{\nu}_{b} +\el^{-2} \big).
\end{equation}
Hence, the requirement of regularity for the most divergent term in \eqref{17VI24.01} implies the asymptotic behavior of the lapse function (cf.\ Eq.~\eqref{notation1}), 
\bea \label{eqlap}
\unph{N}^{-2}=\el^{-2}-\Om \unph f.
\eea

Although the analysis could be continued with a non-vanishing stress-energy tensor satisfying suitable fall-off conditions, we will consider only a vacuum case and assume $T_{ab}=0$ throughout the paper.
Then \eqref{17VI24.01} becomes,
\bea \label{5VII20.3}
    \Omega \tilde{S}_{ab} +2 \tilde{\nabla}_{a} \tilde{\nabla}_{b}\Omega -  \tilde{f} \ \tilde{g} _{ab}
    =0.
\eea
For convenience, we also write down $\Lambda>0$-vacuum Einstein equation in terms of unphysical Ricci tensor,
\begin{equation}\label{unphEFE}
  \unph{R}_{ab} = -\frac{2}{\Om} \unph{\nabla}_a \unph{\nabla}_b \Om - \frac{\unph{g}^ {cd}\unph{\nabla}_c \unph{\nabla}_d \Om }{\Om} \unph{g}_{ab}  + \frac{3}{\Om^2} (\unph{g}^{cd}\unph{\nabla}_c \Om \unph{\nabla}_d \Om  + \el^{-2})\unph{g}_{ab} .
\end{equation}
{We shall refer to \eqref{unphEFE} as conformal Einstein field equations.}

Throughout our paper, we consider that $\tilde f$ has, in general, a smooth non-vanishing contribution on the boundary, i.e. $\tilde{f}\sim \mathcal{O}(1)$. 
However, we note that the conformal Einstein equation \eqref{5VII20.3} is invariant under a rescaling freedom, $\Omega \to \omega \Omega$ and $\unph g \to \omega^2\unph g$. This rescaling of the conformal factor is a gauge freedom. One can use this gauge freedom to set $\tilde {\nabla}_{a} \tilde{\nu}_{b}=0$ on $\scri$ \cite{Geroch:1977jn, Ashtekar:1999jx, Fernandez-Alvarez:2021yog}. This implies $\tilde{f}\sim \mathcal{O}(\Omega)$, and, as a consequence, $\unph N^{-2} = \el^{-2} + \mathcal{O}(\Om^{2})$.

To introduce the perturbation, we consider a one-parameter family of 
physical metrics, $g_{ab}(\epsilon)$ which is differentiable with respect 
to $\epsilon$ at $\epsilon =0$. The first-order perturbation of the physical 
metric is given by $\delta g_{ab}:= \frac{d g_{ab}(\epsilon)}
{d\epsilon}|_{\epsilon=0}$. We assume that the conformal rescaling is independent of the perturbation procedure, i.e., the foliation $\Omega=\const$ does not change, and $\delta \Omega=0$, $\delta \nu_{a}=0$. The linearized Einstein equation for $\delta 
g_{ab}$, expressed in terms of corresponding unphysical perturbation 
$\delta\tilde{g}_{ab}=\Omega^{2} \delta g_{ab}$, is given by
\bea \nn
\hspace*{-.2cm}\tilde{\square} \delta \tilde{g}_{ab}\hspace*{-.3cm}&=& \hspace*{-.3cm}2 \tilde{\nabla}_{(a} \tilde{\nabla}^{m} \delta \tilde {g}_{b)m} - \tilde{\nabla}_{a} \tilde{\nabla}_{b} \delta \tilde{g}_{m}^{m}+ 2 \tilde{R}{^{m}{}_{a}{}^{n}{}_{b}} \delta\tilde{g}{}_{mn}+2 \tilde{R}_{(a}{}^{m} \delta \tilde{g}_{b)m}+\tilde{R}^{mn}\delta \tilde{g}_{mn} \tilde{g}_{ab} 
-\frac{\tilde{R}}{3}(\delta \tilde{g}_{ab}+\frac{1}{2} \delta \tilde{g}_{m}{}^{m}\tilde{g}_{ab})
\\ \nn
&-&2 \Omega^{-1} \tilde{\nu}^{m}\bigg(2\tilde{\nabla}_{(a}\delta \tilde{g}_{b)m}-\tilde{\nabla}_{m}\delta \tilde{g}_{ab}\bigg)
-\Omega^{-1}\bigg(2\tilde{\nu}^{m}\tilde{\nabla}^{n}\delta \tilde{g}_{mn}-\tilde{\nu}^{m}\tilde{\nabla}_{m}  \delta \tilde{g}_{n}{}^{n}\bigg)\tilde{g}_{ab}\\ \label{03.VI.21.1}
&-&2 \Omega^{-2}\bigg(\el^{-2}+\tilde{\nu}_{c}\tilde{\nu}^{c}\bigg) \bigg(\delta \tilde{g}_{ab}+\frac{1}{2} \delta \tilde{g}_{m}{}^{m} \tilde{g}_{ab}\bigg) 
+6\Omega^{-2} \tilde{g}_{ab} \delta \tilde{g}_{cd} \tilde{\nu}^{c}\tilde{\nu}^{d}, \label{8XI22.01}
\eea 
where $\delta \tilde{g}_{m}{}^{m}:= \delta \tilde{g}_{ma}\tilde{g}^{ma}$.\footnote{Beware that we also use the standard convention $\delta\unph{g}^{ab}\equiv\delta g^{\!-1\,ab} = -g^{ac}g^{bd}\delta g_{cd}$ which is however in a conflict with the standard convention for raising indices.}

We will assume a generic fall-off for gravitational perturbation as,
\bea \label{5IX24.01}
\delta \tilde{g}_{ab}= \Omega^{\alpha}\, \tilde{\tau}_{ab},\quad 
\tilde{\tau}_{ab}\, \tilde{\nu}^{{b}}= \Omega^{\beta}\tilde{\tau}_{a}, \quad 
\tilde{\tau}_{a}\, \tilde{\nu}^{a}= \Omega^{\gamma} \tilde{\tau},
 \label{5IX24.02}
\eea
with $\alpha \geq 0, \beta \geq 0, \gamma \geq 0$, and shall denote the trace of the $\unph \tau_{ab}$ field as,
\begin{equation}
    \tilde{\phi}:=\tilde{g}^{ab}\tilde{\tau}_{ab}.
\end{equation}
The fall-off condition introduced in \eqref{5IX24.01}, can also be written as 
\bea 
\delta \tilde{g}_{ab}= \Omega^{\alpha}\, \tilde{\tau}_{ab},\quad 
\delta \tilde{g}_{ab}\, \tilde{\nu}^{{b}}= \Omega^{\alpha+\beta} \, \tilde{\tau}_{a}, \quad 
\delta \tilde{g}_{ab}\, \tilde{\nu}^{a}\, \tilde{\nu}^{b}= \Omega^{\alpha+\beta+\gamma} \,\tilde{\tau}.
\eea
This choice of ansatz is inspired by the analysis of Geroch and Xanthopoulos \cite{Geroch:1978ur}. For clarity, we present the decomposition of the perturbations of the metric \eqref{notation2},
\bea
 \delta \tilde{g}= 
 \begin{bmatrix}
\tilde{N}^{-4}\Omega^{\alpha+ \beta+ \gamma} \, \tilde{\tau} & - \tilde{N}^{-2}\Omega^{\alpha+ \beta} \, \tilde{\tau}_{i} \\
- \tilde{N}^{-2} \Omega^{\alpha+ \beta} \, \tilde{\tau}_{i} & \Omega^{\alpha} \, \tilde{\tau}_{ij}
\end{bmatrix}\,.
\eea

We assume that $\tilde{\tau}_{ab}, \tilde{\tau}_{a}, \tilde{\tau}$ are regular\footnote{{From the perspective of PDE theory, the regularity condition (interpreted as smoothness of the fields) can, in principle, be relaxed to require only an optimal degree of differentiability. However, we will not explore this issue in detail and will simply assume that all fields involved are sufficiently differentiable. }} of the order $\mathcal{O}(1)$ on the conformal boundary~$\scri$. Our goal is to find a suitable choice of $\alpha, \beta, \gamma$ so that the linearized conformal Einstein equation with a positive cosmological constant becomes regular. In terms of these variables, the linearized Einstein equation becomes,
\begin{align}  \nonumber
\tilde{\square}\tilde{\tau}_{ab} &= 2 \tilde{\nabla}_{(a} \tilde{\nabla}^{m} \tilde{\tau}_{b)m} -2 \tilde{R}_{ambn} \tilde{\tau}^{mn}-\frac{\alpha \tilde{\phi}}{2} {\tilde{R}_{ab}}
-\frac{\alpha \tilde{\phi}}{12} \tilde{R} \tilde{g}_{ab} - \tilde{\nabla}_{a} \tilde{\nabla}_{b} \tilde{\phi}+\frac{\alpha}{6} \tilde{R} \tilde{\tau}_{ab} \\ \nonumber
& +\Omega^{-2} \bigg(\el^{-2} \alpha(\alpha -3) \tilde{\tau}_{ab}- \alpha (\alpha -1) \tilde{\phi} \tilde{\nu}_{a} \tilde{\nu}_{b} - \el^{-2} \alpha \tilde{\phi} \tilde{g}_{ab}\bigg) \\ \nonumber
&+\Omega^{-1} \bigg(2\alpha\tilde{\nu}_{(a}\tilde{\nabla}^{m}\tilde{\tau}_{b)m}- 2\alpha \tilde{\nu}_{(a}\tilde{\nabla}_{b)} \tilde{\phi}-\alpha (\alpha -1){\tilde{f}} \tilde{\tau}_{ab}
+ 2(1-\alpha) \tilde{\nu}^{m}\tilde{\nabla}_{m} \tilde{\tau}_{ab}+ \tilde{g}_{ab}\tilde{\nu}^{m} \tilde{\nabla}_{m} \tilde{\phi}\bigg) \\  \nn
&+ 2\Omega^{\beta -2} [(\alpha+\beta)(\alpha -2) -\alpha] \tilde{\nu}_{(a} \tilde{\tau}_{b)}
+ 2 \Omega^{\beta -1} [(\alpha -2) \tilde{\nabla}_{(a}\tilde{\tau}_{b)}-\tilde{g}_{ab}\tilde{\nabla}^{m}\tilde{\tau}_{m}] \\
&+ 2\Omega^{\beta+ \gamma -2}[3-(\alpha+\beta)] \tilde{g}_{ab} \tilde{\tau}. \label{21X23.01}
\end{align}

We now attempt to regularize equation \eqref{21X23.01}.  By ``regularize'' we mean a choice of parameters 
$\alpha,\beta,\gamma$ such that factors multiplying the negative powers of $\Om$ in \eqref{21X23.01} cancel 
out or can be gauged away, so that the resulting equation is regular at $\{ \Om = 0 \}$. Note that one can 
always divide by the most negative power of $\Om$ and get a regular equation at $\{ \Om = 0 \}$, but then the 
principal symbol of the equation vanishes at that hypersurface and therefore fails to be hyperbolic. For 
$\Lambda =0$ such choice was shown by Geroch and Xanthopoulos \cite{Geroch:1978ur}  to be $\alpha=1=\beta, 
\gamma=0$ (we review this in Section~\ref{seclambdazero}). As we shall next see, for non-zero $\Lambda$ this ansatz does not make 
the leading order divergent term (of order $\Omega^{-2}$) vanish. Moreover, this term contains $\unph 
\tau_{ab}$ and therefore there is no chance that this can be gauged away. 
This is a remarkable difference entirely due to the presence of a cosmological constant.

Let us start by examining the leading order divergent term at $\Om = 0$ for non-zero $\Lambda$. From equation \eqref{21X23.01} this is
\begin{equation}\label{leadingdivergence}
\begin{split}
   &\Omega^{-2} \bigg(\el^{-2} \alpha(\alpha -3) \tilde{\tau}_{ab}- \alpha (\alpha -1) \tilde{\phi} \tilde{\nu}_{a} \tilde{\nu}_{b} -  \el^{-2} \alpha \tilde{\phi} \tilde{g}_{ab}\bigg)\\
   &\qquad+2\Omega^{\beta -2} \Bigl((\alpha+\beta)(\alpha -2) -\alpha\Bigr) \tilde{\nu}_{(a} \tilde{\tau}_{b)}
   +2\Omega^{\beta+ \gamma -2}\Bigl(3-(\alpha+\beta)\Bigr) \tilde{g}_{ab} \tilde{\tau},
\end{split}
\end{equation}
where $\Om^{\beta-2}$ and $\Om^{\beta+\gamma-2}$ terms contribute only if $\beta=0$ and $\beta=\gamma=0$, respectively. We need \eqref{leadingdivergence} to vanish or {at least, be such that it can be}  gauged away. Thus, it cannot contain the complete field $\unph \tau_{ab}$, but there may only appear certain components, such as a trace or a divergence of it.  Also note that the term $\unph \tau_{ab}$ cannot be removed using the terms of $\Om^{\beta -2}$ and $\Om^{\beta + \gamma -2}$ either. Thus, in the case of non-zero $\Lambda$ one is forced to choose $\alpha = 0$.\footnote{The case $\alpha = 3$ faces problems at $\Om^{-1}$ order and leads to vanishing symplectic structure. Therefore we do not consider this option here.}  This is a central difference between the zero and non-zero $\Lambda$ cases. 

Therefore, the highest order divergent terms $\Omega^{-2}$ vanishes for $\alpha=0$. A possible choice for the regularization of the $\Omega^{-1}$ term in \eqref{21X23.01} is $\alpha=0,\, \beta=0, \gamma=1$. The regularity condition then reads
\bea \label{10X2024.01}
2 \tilde{\nu}^{m}\tilde{\nabla}_{m} \tilde{\tau}_{ab}+\tilde{g}_{ab} \tilde{\nu}^{m}\tilde{\nabla}_{m} \tilde{\phi}-4\tilde{\nabla}_{(a}\tilde{\tau}_{b)}-2\tilde{g}_{ab} \tilde{\nabla}^{m} \tilde{\tau}_{m}+ 6 \tilde{g}_{ab} \tilde{\tau}=\mathcal{O}(\Omega).
\eea
Taking the trace of this equation, we obtain
\bea \label{10X2024.02}
\tilde{\nu}^{m}\tilde{\nabla}_{m}\tilde{\phi}-2\tilde{\nabla}^{m}\tilde{\tau}_{m}+4\tilde{\tau}=\mathcal{O}(\Omega).    
\eea
Substracting \eqref{10X2024.02} from \eqref{10X2024.01} multiplied by $\tilde{g}_{ab}$, gives
\bea  \label{25XI24.01}
-2\tilde{\nabla}_{(a}\tilde{\tau}_{b)}+\tilde{\nu}^{m} \tilde{\nabla}_{m} \tilde{\tau}_{ab}+\tilde{g}_{ab} \tilde{\tau} = \mathcal{O}(\Omega).
\eea
Note that this condition is fully equivalent to \eqref{10X2024.01}.
{Similar to the discussion of the $\Om^{-2}$ order above, this term has no chance to be canceled out by a choice of gauge, because that would imply a too strong condition on  $\tilde{\nu}^{m} \tilde{\nabla}_{m} \tilde{\tau}_{ab}$. Thus we must require that $\unph\tau_a$ satisfies the constraint 
\bea  
\label{constr}
-2\tilde{\nabla}_{(a}\tilde{\tau}_{b)}+\tilde{\nu}^{m} \tilde{\nabla}_{m} \tilde{\tau}_{ab}+\tilde{g}_{ab} \tilde{\tau} = \Om \aux_{ab}
\eea
with some regular field $\aux_{ab}$. Equation \eqref{21X23.01} then reads 
\bea  
\tilde{\square}\tilde{\tau}_{ab} &=& 2 \tilde{\nabla}_{(a} \tilde{\nabla}^{m} \tilde{\tau}_{b)m} -2 \tilde{R}_{ambn} \tilde{\tau}^{mn} - \tilde{\nabla}_{a} \tilde{\nabla}_{b} \tilde{\phi} + \aux_{ab} \label{regularized}.
\eea
We now have a smooth equation at $\Om = 0$. However, the system is incomplete as there is no dynamical equation for $\aux_{ab}$. In this situation, one needs to check if the core equation \eqref{regularized} propagates the field $\aux_{ab}$ too. There is no well-defined prescription on how to do this, although this usually involves taking derivatives of \eqref{regularized} and using again definition \eqref{constr} and \eqref{regularized} to try to obtain a new propagation equation for  $\aux_{ab}$. Unfortunately, this seems to introduce new divergent terms that must be eliminated by defining additional auxiliary fields, leading to an infinite process of solving the equations order-by-order in $\Omega$. 

In summary, we have found that the choice of parameters $\alpha = \beta = 0$ and $\gamma = 1$ leads to a constraint of the form \eqref{25XI24.01}. Such a constraint does not appear to be removable by any other choice of parameters. Since the choice $\alpha = \beta = 0$ and $\gamma = 1$ is minimal in the sense that it includes all other cases as subcases, we shall refer to this condition as the minimal fall-off behaviour of the linearized fields in de Sitter spacetimes.

If we wish to extend this fall-off behaviour up to the conformal boundary $\scri$, additional results are required to guarantee the extendibility of linear fields to $\scri$. As the well-posedness of the full system of equations lies beyond the scope of this paper, we instead rely on the nonlinear stability results derived from Friedrich's conformal field equations with $\Lambda > 0$ \cite{Friedrich:1986qfi}. This implies linear stability of the fields and thus extedibility of the linear fields up to $\scri$  (see subsection \ref{linstab} of Appendix \ref{apphyperbolic}). With this result at hand, the condition $\alpha = \beta = 0$, $\gamma = 1$ does indeed yield the general fall-off behaviour of linear fields with $\Lambda > 0$.
 In the remainder of this paper, we will assume the minimal fall-off behaviour and examine its implications.


We also emphasize that the regularization condition in \eqref{25XI24.01} can also be understood as the regularization of $\tilde{f}=\Omega^{-1}(\tilde{g}^{cd}\tilde{\nu}_{c}\tilde{\nu}_{d}+\el^{-2})$ under the first order perturbation. Indeed, thanks to $\delta \Omega =0=\delta \tilde{\nu}_{a}$ we obtain
\bea 
-\delta \tilde{g}_{cd}\tilde{\nu}^{c}\tilde{\nu}^{d}= \Omega \delta \tilde{f}.
\eea
Hence $\delta \tilde{g}_{ab}\tilde{\nu}^{a}\tilde{\nu}^{b}=\mathcal{O}(\Omega)$, which corresponds to $\alpha+ \beta+ \gamma=1$. To match with the notation in \eqref{5IX24.01}, we obtain $\delta \tilde{f}=-\tilde{\tau}$. From \eqref{5VII20.3}, we also have 
\bea \label{21VII24.01}
\tilde{\nabla}_{a} \tilde{\nu}_{b} \big|_{\scri}=\frac{\tilde{f}}{2} \tilde{g}_{ab}.
\eea
Under the linearized perturbation, this equation transforms as
\bea \nonumber
&&-\frac{\tilde{\nu}^{c}}{2} \bigg(\tilde{\nabla}_{a}\delta \tilde{g}_{bc}+ \tilde{\nabla}_{b}\delta \tilde{g}_{ac}-\tilde{\nabla}_{c}\delta \tilde{g}_{ab}\bigg)= \frac{\delta \tilde{f}}{2} \tilde{g}_{ab}+\frac{\tilde{f}}{2} \delta \tilde{g}_{ab}+ \mathcal{O}(\Omega),\\
&&\implies \tilde{\nabla}_{a}(\tilde{\nu}^{c}\delta \tilde{g}_{bc})+ \tilde{\nabla}_{b}(\tilde{\nu}^{c}\delta \tilde{g}_{ac})-\tilde{\nu}^{c}\tilde{\nabla}_{c}\delta \tilde{g}_{ab}=-{\delta \tilde{f}}\tilde{g}_{ab}+\mathcal{O}(\Omega).\label{31III25.01}
\eea
With the `minimal' fall-off condition, $\alpha=0=\beta, \gamma=1$, \eqref{31III25.01} reduces to \eqref{25XI24.01}.
We emphasize that this constraint equation is trivially satisfied for Geroch and Xanthopoulos' anstaz $\alpha=1=\beta, \gamma=0$, in the context of asymptotically flat space-times.

\section{Regularized conformal Einstein equation for vanishing cosmological constant}\label{seclambdazero}

We next review the regularization of \eqref{21X23.01} carried out by Geroch and Xanthopoulos for the $\Lambda = 0$ case in \cite{Geroch:1978ur}. {A priori, $\Lambda=0$ case of the equation \eqref{21X23.01} is not clear}. For this case it is convenient to write down \eqref{21X23.01} in a different form,
\bea  \nonumber
\tilde{\square}\tilde{\tau}_{ab} &=& 2 \tilde{\nabla}_{(a} \tilde{\nabla}^{m} \tilde{\tau}_{b)m} -2 \tilde{R}_{ambn} \tilde{\tau}^{mn}+\tilde{R}^{mn} \tilde{\tau}_{mn} \tilde{g}_{ab}+\frac{\alpha \tilde{\phi}}{2} {\tilde{R}_{ab}}
-\frac{\tilde{R}}{6} \tilde{\phi} \tilde{g}_{ab} (1+\frac{\alpha}{2})- \tilde{\nabla}_{a} \tilde{\nabla}_{b} \tilde{\phi}+\frac{\alpha}{6} \tilde{R} \tilde{\tau}_{ab} \\ \nonumber
& +& \Omega^{-2}  \alpha (1-\alpha) \tilde{\phi} \tilde{\nu}_{a} \tilde{\nu}_{b}  
+\Omega^{-1} \bigg(2\alpha\tilde{\nu}_{(a}\tilde{\nabla}^{m}\tilde{\tau}_{b)}m- 2\alpha\tilde{\nu}_{(a}\tilde{\nabla}_{b)} \tilde{\phi}-\alpha (\alpha -1)\tilde{f}\tilde{\tau}_{ab}\\ \nonumber
&+& 2(1-\alpha) \tilde{\nu}^{m}\tilde{\nabla}_{m} \tilde{\tau}_{ab}+ \tilde{g}_{ab}\tilde{\nu}^{m} \tilde{\nabla}_{m} \tilde{\phi}-2 \tilde{g}_{ab} \tilde{\nu}^{m} \tilde{\nabla}^{n} \tilde{\tau}_{mn} -\tilde{\phi} \tilde{f} \tilde{g}_{ab}(1-\frac{\alpha}{2})\bigg) \\ 
\label{23X23.01}
&+& 2\Omega^{\beta -2} [(\alpha+\beta)(\alpha -2) -\alpha] \tilde{\nu}_{(a} \tilde{\tau}_{b)}
+ 2 \Omega^{\beta -1} [(\alpha -2) \tilde{\nabla}_{(a}\tilde{\tau}_{b)}] 
+ 2\Omega^{\beta+ \gamma -2}(3-\alpha) \tilde{g}_{ab} \tilde{\tau}.
\eea
In deriving equation \eqref{23X23.01}, we have used the identity 
\bea \label{7III24.01}
\Omega^{-1} \tilde{\nabla}_{a} \tilde{\nu}_{b} = - \frac{\tilde{R}_{ab}}{2}+\frac{1}{12} \tilde{R} \tilde{g}_{ab}+\Omega^{-1} \frac{\tilde{f}}{2} \tilde{g}_{ab},
\eea
which follows by writing the trace components of \eqref{unphEFE} in terms of $R$.
The suitable choice of parameters for $\Lambda = 0$ made in \cite{Geroch:1978ur}
is $\alpha=1=\beta$, and $\gamma=0$. Then \eqref{23X23.01} becomes
\bea \nonumber
\tilde{\square}\tilde{\tau}_{ab}&=& 2\tilde{\nabla}_{(a}\tilde{y}_{b)}-2\tilde{R}_{ambn} \tilde{\tau}^{mn}+\tilde{R}^{mn} \tilde{\tau}_{mn} \tilde{g}_{ab}+\frac{\tilde{R}}{6}  \tilde{\tau}_{ab} + \tilde{\nabla}_{a} \tilde{\nabla}_{b} \tilde{\phi}-\frac{\tilde{R}}{4} \tilde{\phi} \tilde{g}_{ab}+\frac{\tilde{\phi}}{2} \tilde{R}_{ab}+ 4 \tilde{\nabla}_{{(a}}\tilde{\tau}_{b)} \\ \label{16VI24.01}
&+&2\Omega^{-1}\bigg(\tilde{\nu}_{(a}\tilde{y}_{b)}-\tilde{g}_{ab} \tilde{\nu}^{m} \tilde{y}_{m} -\tilde{g}_{ab}(\frac{1}{2} \tilde{\nu}^{c}\tilde{\nabla}_{c}\tilde{\phi}+\tilde{\nu}^{c}\tilde{\tau}_{c}+\frac{\tilde{f}}{4} \tilde{\phi})\bigg),
\eea
where we have used the definition
\bea 
\tilde{y}_{{a}}&:=& \tilde{\nabla}^{m} \tilde{\tau}_{ma}-\tilde{\nabla}_{a} \tilde{\phi}-3\tilde{\tau}_{a}.
\eea
\ifx
Introducing new variables
\begin{eqnarray}
\tilde{y}_{b} &:=& \tilde{\nabla}^{m} \tilde{\tau}_{mb} -\tilde{\nabla}_{b} \tilde{\tau} -3 \tilde{\tau}_{b},\\ 
\tilde{\sigma}&:=& \Omega^{-1} (\tilde{n}^{m} \tilde{\tau}_{m}+\frac{1}{2} \tilde{n}^{m}\tilde{\nabla}_{m} \tilde{\tau}+\frac{1}{4} \tilde{f}\tilde{\phi}),
\end{eqnarray}
\fi
Introducing the auxiliary field variable 
\bea 
\tilde{\sigma}&:=& \Omega^{-1}(\tilde{n}^{a}\tilde{\tau}_{a}+\frac{1}{2} \tilde{n}^{a}\tilde{\nabla}_{a}\tilde{\phi}+\frac{1}{4} \tilde{f}\tilde{\phi}),
\eea
and using the identity between Riemann tensor and Weyl tensor,  we  rewrite equation \eqref{16VI24.01} as
\begin{eqnarray} \nonumber
\tilde{\square} \tilde{\tau}_{ab}&=& 2 \tilde{\nabla}_{(a}\tilde{y}_{b)} +\tilde{\nabla}_{a} \tilde{\nabla}_{b} \tilde{\phi} +4\tilde{\nabla}_{(a} \tilde{\tau}_{b)} - 2 \tilde{C}_{ambn} \tilde{\tau}^{mn} -\frac{1}{6} \tilde{R} \tilde{\tau}_{ab}+\frac{1}{12} \tilde{R} \tilde{\phi} \tilde{g}_{ab} -\frac{1}{2} \tilde{R}_{ab} \tilde{\phi} +2 \tilde{\tau}^{m}_{(a}\tilde{R}_{b)m} - 2 \tilde{\sigma} \tilde{g}_{ab}\\ \label{23X23.03}
&+&2\Omega^{-1} (\tilde{\nu}_{(a}\tilde{y}_{b)}-\tilde{g}_{ab} \tilde{\nu}^{m} \tilde{y}_{m})
\end{eqnarray}
For $\Lambda=0$ case, we can choose the following gauge conditions (cf. \cite{Geroch:1978ur})
\bea \label{6VI24.01}
\tilde{y}_{a}:=\tilde{\nabla}^{b} \tilde{\tau}_{ab}-\tilde{\nabla}_{a}\tilde{\phi}-3\tilde{\tau}_{a}&=&0,\\
\label{6VI24.02}
(\tilde{n}^{a}\tilde{\nabla}_{a}+\frac{1}{6}\Omega \tilde{R}+\frac{3}{2}\tilde{f}) \tilde{\nabla}^{2}\tilde{\phi}&=&\frac{1}{12} \tilde{R} \tilde{f} \tilde{\phi} -\frac{1}{2} \tilde{\phi} \tilde{\nabla}^{2} \tilde{f} -\frac{1}{3} \tilde{R} \tilde{n}^{a} \tilde{\tau}_{a}+4 \Omega^{-1} \tilde{C}_{ambn} \tilde{\tau}^{mn} \tilde{n}^{a} \tilde{n}^{b}.
\eea
Therefore, from \eqref{6VI24.01}, equation \eqref{23X23.03} reduces to 
\begin{align} 
\tilde{\square} \tilde{\tau}_{ab}&=& \tilde{\nabla}_{a} \tilde{\nabla}_{b} \tilde{\phi} +4\tilde{\nabla}_{(a} \tilde{\tau}_{b)} - 2 \tilde{C}_{ambn} \tilde{\tau}^{mn} -\frac{1}{6} \tilde{R} \tilde{\tau}_{ab}+\frac{1}{12} \tilde{R} \tilde{\phi} \tilde{g}_{ab} -\frac{1}{2} \tilde{R}_{ab} \tilde{\phi} +2 \tilde{\tau}^{m}_{(a}\tilde{R}_{b)m} - 2 \tilde{\sigma} \tilde{g}_{ab}.\label{reglambda0}
\end{align}
We have now removed all divergent terms, but we still  need to introduce some dynamical equation for the {other components of the} fields $\sigma$ and $\tilde \tau_a$. Contracting \eqref{reglambda0} with $\unph \nu^b$ leads (after doing some algebra) to 
\begin{align}
    \tilde{\square} \tilde{\tau}_{a} &= 2 \tilde{\nabla}_{a} \tilde{\sigma} + \frac{1}{2} \tilde{R}_{am} \tilde{\nabla}^{m} \tilde{\phi} + \frac{1}{12} \tilde{R} \tilde{\nabla}_{a} \tilde{\phi} - \tilde{R}^{mn} \tilde{\nabla}_{m} \tilde{\tau}_{an} - \frac{1}{3} \tilde{\tau}_{ab} \tilde{\nabla}^{b} \tilde{R} \notag \\
    &\quad + 2 \tilde{\tau}^{mn} \tilde{\nabla}_{[m} \tilde{R}_{a]n} + 2 \tilde{\tau}^{m} \tilde{R}_{ma} + \frac{1}{2} \tilde{R} \tilde{\tau}_{a} + \frac{1}{2} \tilde{\phi} \tilde{\nabla}_{a} \tilde{R}.\label{eqtau}
\end{align}
Similarly, a second contraction with $\unph \nu^a$, together with gauge condition \eqref{6VI24.02} gives
\begin{align}
    \tilde{\square} \tilde{\sigma} &= - \frac{1}{2} \tilde{R}^{mn} \tilde{\nabla}_{m} \tilde{\nabla}_{n} \tilde{\phi} - 2 \tilde{R}^{mn} \tilde{\nabla}_{m} \tilde{\tau}_{n} - \frac{1}{12} (\tilde{\nabla}_{m} \tilde{R}) (\tilde{\nabla}^{m} \tilde{\phi}) + \tilde{R} \tilde{\sigma} \notag \\
    &\quad + \frac{1}{72} \tilde{R}^2 \tilde{\phi} - \frac{1}{2} \tilde{\tau}_{ab} \tilde{R}^{am} \tilde{R}^{b}_{\ m} - \frac{1}{3} \tilde{\tau}^{m} \tilde{\nabla}_{m} \tilde{R}.\label{eqsigma}
\end{align}
The system of equations  \eqref{reglambda0}, \eqref{eqtau} and \eqref{eqsigma},  supplemented with 
the gauge conditions \eqref{6VI24.01}, \eqref{6VI24.02} with some additional variables satisfying 
wave equations and transport equations, turns out to give a hyperbolic PDE 
system \cite{Geroch:1978ur}. A higher (even) dimensional generalisation of this formalism with different choice of gauge 
and field variables is also available in \cite{Hollands:2003ie}.

\section{Gauge transformation of perturbation} \label{Sec:Gauge}
Linearized perturbations are defined up to gauge transformations corresponding to infinitesimal diffeomorphisms generated by a vector field $\xi^a$
\bea
\delta g_{ab} \mapsto \delta g'_{ab}=\delta g_{ab} + \mathcal{L}_\xi g_{ab}=\delta g_{ab} + 2\nabla_{(a} \xi_{b)}\label{eqgaugetrans},
\eea
where $ \mathcal{L}_\xi$ stands for the Lie derivative with respect to $\xi^a$ and $\xi_a = \phy g_{ab} \xi^b$. Consider the generator of an infinitesimal diffeomorphism of the form
\begin{equation}
    \xi^a =  \zeta_\perp \phy n^a + \zeta^a
\end{equation}
where $\zeta_\perp$ is a scalar field and $\zeta^a$ satisfies $\zeta^a = \xi^b \spacemet^a{}_b$, both $\mathcal{O}(1)$ with respect to $\Om$.  For linear fields $\delta \phy g_{ab}$ admitting a conformal extension, $\delta \unph g_{ab} = \Om^2 \delta \phy g_{ab}$, adding a gauge transformation may spoil the conformal extendability property if the normal component of $\xi^a$ does not vanish at $\scri$. As we shall see next, this is precisely the behaviour of $\xi^a$ as prescribed. Note  that although we ask  $\zeta_\perp = \mathcal{O}(1)$, this is the normal component wrt $n^a$, which is not defined at $\{\Om = 0\}$, so we must first express $\xi^a$ in terms of objects extendable to $\{ \Om = 0\}$. 

Since diffeomorphisms are metric independent transformations and its generators are naturally the vector fields $\xi^a$, we do not rescale them when considering unphysical gauge transformations. Namely, we shall consider unphysical gauge transformations generated by $\unph \xi^a = \xi^a$. In order to write $\unph \xi$ it in terms of unphysical quantities observe
\begin{equation}
    \unph \zeta_\perp = \unph \xi^a \unph n_a = \xi^a \unph n_a = \zeta_\perp  n^a \unph n_a = \Om \zeta_\perp,
    \quad \unph \zeta^a = \unph \xi^b \unph\spacemet^a{}_b = \xi^b \spacemet^a{}_b = \zeta^a.
\end{equation}
Therefore,
\begin{equation}\label{equnphgauge}
    \unph\xi^a = \Om \zeta_\perp \unph n^a + \unph \zeta^a.
\end{equation}
Note that $\unit^a$ is a regular vector at $\scri$ and by hypothesis $\zeta_\perp = \mathcal{O}(1)$ and $\unph \zeta^a =  \zeta^a = \mathcal{O}(1)$. Hence $\unph\xi^a$ is extendable to $\scri$. Since its normal component vanishes for $\Om = 0$, this means that $\unph \xi$ becomes tangent at $\scri$. Thus, the diffeomorphisms generated by vectors of the form \eqref{equnphgauge} preserve the locus of $\scri$. \footnote{We note that for vanishing cosmological constant, when splitting the vector field $\xi^a$ tangential to null $\scri$ in similar way as in \eqref{equnphgauge} we have to use null (non-normalized) normal $\unph \nu^{a}:=\unph g^{ab}d_{b}\Omega$, namely $\unph \xi^{a}=\unph \zeta\tilde{\nu}^{a}+ \unph \zeta^{a}$. Therefore, we do not need any special behaviour of $\unph \zeta$.}

We may now write the gauge transformation \eqref{eqgaugetrans} for the unphysical linear fields 
\begin{align}
    \delta \unph g'_{ab} = \Om^2 \delta g'_{ab} & = \Om^{2} ( \delta g_{ab} + \mathcal{L}_\xi g_{ab}) =  \delta \unph g_{ab} + \Om^{2} \mathcal{L}_{\unph \xi} (\Om^{-2}\unph g_{ab}) = \delta \unph g_{ab} - 2\Om^{-1} \unph \xi(\Om) \unph g_{ab} + 2 \unph\nabla_{(a} \unph \xi_{b)} \nonumber\\
    & = \delta \unph g_{ab} + 2 \lap^{-1} \zeta_\perp \unph g_{ab} + 2 \zeta_\perp \lap^{-1}  \unph n_{(a} \unph n_{b)} + {2 \Om \zeta_\perp \unph \nabla_{(a}\unph n_{b)}}  +2 \Om  \unph n_{(a} \unph\nabla_{b)}\zeta_\perp + 2 \unph\nabla_{(a} \unph \zeta_{b)} \nonumber\\
    & = \delta \unph g_{ab} + 2 \lap^{-1} \zeta_\perp \unph \spacemet_{ab}  +{2 \Om \zeta_\perp \unph \nabla_{(a}\unph n_{b)}}+2 \Om  \unph n_{(a} \unph\nabla_{b)}\zeta_\perp + 2 \unph\nabla_{(a} \unph \zeta_{b)}\label{unphgauge}.
\end{align}
 Now it is a matter of direct calculation to check that \eqref{unphgauge} respects  the `minimal' fall-off behaviour of the linearized field. For $\delta \unph g'_{ab}$ and $ \unph n^b \delta \unph g'_{ab}$ it is immediate that
 \begin{align}
     \delta \unph g'_{ab} & = \delta \unph g_{ab} + 2 \lap^{-1} \zeta_\perp \unph \spacemet_{ab} +{2 \Om \zeta_\perp \unph \nabla_{(a}\unph n_{b)}} +2 \Om  \unph n_{(a} \unph\nabla_{b)}\zeta_\perp + 2 \unph\nabla_{(a} \unph \zeta_{b)} = \mathcal{O}(1),\label{foffdelta} \\
    \unph n^b \delta \unph g'_{ab} & =  \unph n^b \delta \unph g_{ab} +{ \Om \zeta_\perp \unph n^b \unph \nabla_{b}\unph n_{a}} - \Om \unph\nabla_a \zeta_\perp  + \Om \unph n_a \unph n^b \unph\nabla_b \zeta_\perp + 2 \unph n^{{b}} \unph\nabla_{(a}\unph \zeta_{b)} = \mathcal{O}(1).\label{foffndelta}
 \end{align}
For $ \unph n^a \unph n^b \delta \unph g'_{ab} $ we have 
\begin{equation}
   \unph n^a \unph n^b \delta \unph g'_{ab}= \unph n^a \unph n^b \delta \unph g_{ab} - 2 \Om \unph n^a \unph\nabla_a \unph \zeta_\perp + 2 \unph n^a \unph n^b \unph\nabla_{(a}\unph \zeta_{b)} = {\unph n^a \unph n^b \delta \unph g_{ab}}- 2 \Om \unph n^a \unph\nabla_a \unph \zeta_\perp -  {2}\unph n^a \unph \zeta^b \unph\nabla_{a}\unph n_{b},
\end{equation}
where we have used that $\unph n^a \unph \zeta_a = 0 $ {and $\unph n^a \unph \nabla_b \unph n_a = 0$}. It follows from 
the Einstein equations (cf. \eqref{unnphEFE1}) that $\unph\nabla_{a}\unph n_{b}$ has $\mathcal{O}(1)$ trace terms 
plus $\mathcal{O}(\Om)$ terms. Thus $ \unph n^a \unph \zeta^b \unph\nabla_{a}\unph n_{b} = \mathcal{O}(\Om)$ and
\begin{equation}
        \unph n^a \unph n^b \delta \unph g'_{ab} = \unph n^a \unph n^b \delta \unph g_{ab} - 2 \Om \unph n^a \unph\nabla_a \unph \zeta_\perp - {2}\unph n^a \unph \zeta^b \unph\nabla_{a}\unph n_{b} = \mathcal{O}(\Om) \label{foffnndelta}
\end{equation}
Taking into account that $\unit^a = \lap \grOm^a$ and that $\lap = \mathcal{O}(1)$ (cf. Eq. \eqref{eqlap}, also Section \ref{seceinstein} below), it is ready to translate \eqref{foffdelta},\eqref{foffndelta} and \eqref{foffnndelta} into conditions on the fields $\{ \tau_{ab},\tau_a, \tau\}$  under change of gauge
\begin{align}
    \tau'_{ab} & = \delta g'_{ab} = \mathcal{O}(1), \label{eqtauabdg}\\ 
    \tau'_{a} & = \grOm^b \delta g'_{ab} = N^{-1} \unit^b \delta g'_{ab}  = \mathcal{O}(1), \label{eqtaubdg}\\ 
     \tau' & = \grOm^a\grOm^b \delta g'_{ab} = N^{-2} \unit^a\unit^b \delta g'_{ab}  = \mathcal{O}(\Om)\label{eqtaudg},
\end{align}
which match the analysis in Section \ref{secregularization}.

Therefore, we have considered the gauge fields of 
the form \eqref{equnphgauge}. These vector fields 
generate rather general class of diffeomorphisms 
of $\unph M$, where the only restriction that 
they must preserve the locus of $\scri$. In 
addition, it turns out that this kind of gauge 
transformations respect the fall-off conditions 
of the linear fields obtained directly from the 
Einstein equation in Section 
\ref{secregularization}.

\section{Gauge degrees of freedom}\label{secdegfreedom}

As mentioned in the introduction, several recent literature \cite{Kolanowski:2021hwo, Anninos:2010zf, Compere:2019bua, Compere:2020lrt, Poole:2021avh} analyze gravitational waves in presence of positive cosmological constant in Fefferman-Graham gauge. The unphysical metric in the Fefferman-Graham gauge is written as
\begin{equation}
    \unph g_{ab} = -\el^{2}\tilde{\nu}_{a}\tilde{\nu}_{b} + \unph \gamma_{ab},
\end{equation}
where $\unph \gamma$ is an object intrinsic to the leaves with $ \Om = \const $. If one assumes a 1-parameter family of metrics $\tilde g(\epsilon)$ sharing this gauge, the linear fields that one obtains are $\delta \unph g_{ab} = \delta \unph \gamma_{ab}$ thus satisfying $\delta \unph g_{ab} \unph n^a = 0$. In the notation of section \ref{secregularization}, this amounts to set $\tau = 0$ and $\tau_a = 0$. Our goal in this section is to verify that this indeed corresponds to a choice of linear gauge, which, moreover, we will show it is independent of the gauge one uses for the background fields. The strategy consists in splitting $\delta g'_{ab}$ into components that yield an evolution problem (via equation \eqref{unphgauge}) for gauge fields $\{\ngen, \unph\tgen^a \}$ which we can control. Then we check that there always exists a choice of gauge fields which yield  $\tau = 0$ and $\tau_a = 0$. This result makes explicit that the gauge degrees of freedom are localized into the normal-tangent and normal-normal components of the linear fields.

First, consider the gauge transformations discussed in the previous section in the following form:
\begin{align}
    \delta \unph g'_{ab}  & = \delta \unph g_{ab} - 2\Om^{-1}  \gen(\Om) \unph g_{ab} + 2 \unph\nabla_{(a} \unph \xi_{b)} 
     = \delta \unph g_{ab} + 2 \lap^{-1} \zeta_\perp \unph g_{ab} + 2 \unph\nabla_{(a} \unph \xi_{b)}
\end{align}
The normal-tangent terms of above equation, with definition $v_c = {\unit^a \unph{\spacemet}^b{}_c \delta \unph{g} }_{ab}$, are
\begin{align}
 \unit^a \unph{\spacemet}^b{}_c \delta \unph{g}'_{ab} & =  \unit^a \unph{\spacemet}^b{}_c \left( \unph{\nabla}_a \gen_b + \unph{\nabla}_b \gen_a \right) +  {\unit^a \unph{\spacemet}^b{}_c \delta \unph{g} }_{ab}
 \nonumber \\ & =  
 \unit^a  \left( \unph{\nabla}_a (\unph{\spacemet}^b{}_c \gen_b) - (\unph{\nabla}_a \unph{\spacemet}^b{}_c) \gen_b  \right) +  \unph{\spacemet}^b{}_c \left( \unph{\nabla}_b (\unit^a \gen_a) - (\unph{\nabla}_b \unit^a) \gen_a  \right) + v_c \nonumber
 \\ & 
 =   \unit^a \unph{\nabla}_a  \unph\tgen_c - \unit^a \unph{\nabla}_a \unph{\spacemet}^b{}_c \left( {\Om \ngen}\unit_b + \unph\tgen_b \right ) - \Om  \unph{\spacemet}^b{}_c \unph{\nabla}_b \ngen - \unph{\spacemet}^b{}_c  \unph{\nabla}_b \unit^a \left( {\Om \ngen}\unit_a + \unph\tgen_a \right ) + v_c \nonumber 
 \\ & = \unit^a \unph{\nabla}_a  \unph\tgen_c - \left( \unit^a \unph{\nabla}_a \unph{\spacemet}^b{}_c + \unph{\spacemet}^a{}_c \unph{\nabla}_a \unit^b \right) \unph\tgen_b \nonumber\\  & ~ - \Om \left( \unph{\spacemet}^b{}_c \unph{\nabla}_b \ngen + \left(  {\unit^a \unit^b} \unph{\nabla}_a \unph{\spacemet}_{bc} + {\unph{\spacemet}^b{}_c \unit_a } \unph{\nabla}_b \unit^a \right) \ngen \right) + v_c
 \nonumber 
 \\ & = \unit^a \unph{\nabla}_a  \unph\tgen_c - \left( \unit^a \unph{\nabla}_a \unph{\spacemet}^b{}_c + \unph{\spacemet}^a{}_c \unph{\nabla}_a \unit^b \right) \unph\tgen_b \nonumber - \Om \left( \unph{\spacemet}^b{}_c \unph{\nabla}_b \ngen + \left(  {\unit^a \unit^b} \unph{\nabla}_a \unph{\spacemet}_{bc} \right) \ngen \right) + v_c,
\end{align}
where for the last equality we have used that $\unit_a  \unph{\nabla}_b \unit^a = 0 $. Using $\K_{ab} = \unph \nabla_a \unit_b$ we may write 
\begin{align}
 \unit^a \unph{\spacemet}^b{}_c \delta \unph{g}'_{ab}  = \unit^a \unph{\nabla}_a  \unph\tgen_c - \left( \unit^a \unph{\nabla}_a \unph{\spacemet}^b{}_c + \unph{\spacemet}^a{}_c \K_a{}^b \right) \unph\tgen_b  - \Om \left( \unph{\spacemet}^b{}_c \unph{\nabla}_b \ngen -   \unph{\spacemet}_{bc} \unit^a  \K_a{}^b   \ngen \right) + v_c, \label{hypngamdelta1}
\end{align}
For the normal-normal components, we use \eqref{unphgauge}
\begin{align}
 \unit^a \unit^b\delta \unph{g}'_{ab} 
 & = \unit^a \unit^b \left( \delta \unph g_{ab} + 2 \lap^{-1} \zeta_\perp \unph \spacemet_{ab}  +2 \Om  \unph n_{(a} \unph\nabla_{b)}\zeta_\perp + 2 \unph\nabla_{(a} \unph \zeta_{b)} \right) \nonumber \\ 
 &  = -2 \Om \unit^b \unph \nabla_b \ngen + 2 \unit^a \unit^b \unph \nabla_a \unph \tgen_b + \unit^a \unit^b \delta g_{ab}  \\ 
 &= -2 \Om \unit^b \unph \nabla_b \ngen -  2 \unit^a \unph \tgen_b \unph \nabla_a \unit^b + \unit^a \unit^b \delta g_{ab} \label{hypnndelta1}
\end{align}
Using the unphysical Einstein equation \eqref{unphEFE} we have 
 \begin{align}
 {\unit^a \unph \tgen^b }\unph \nabla_a \unit^b = -\Om \lap\frac{\unph{R}_{ab}}{2} \unit^a \unph \tgen^b. \nonumber
\end{align}
In addition, since $\unit^a \unit^b \delta g_{ab} = O(\Om)$ (cf. Section \ref{secregularization}), we may write $\unit^a \unit^b \delta g_{ab} = \Om u$ for a field $u$ regular at $\scri$ and thus equation \eqref{hypnndelta1} yields
\begin{align}
 \unit^a \unit^b\delta \unph{g}'_{ab} = -2 \Om \unit^b \unph \nabla_b \ngen + \Om \lap{\unph{R}_{ab}} \unit^a \unph \tgen^b+ \Om u .
\end{align}
We will next show that there always exists a gauge choice such that
\begin{equation}\label{eqalinFGgauge}
    \unit^a \unph \spacemet^b{}_c \delta g'_{ab} = 0,\qquad  \unit^a \unit^b \delta g'_{ab} = 0.
\end{equation}
From equations \eqref{hypngamdelta1} and \eqref{hypnndelta1} this amouts to finding gauge fields $\{\ngen,\unph \tgen^a \}$ solving the following PDE system
\begin{align}
\unit^a \unph{\nabla}_a  \unph\tgen_c - \Om  \unph{\spacemet}^b{}_c \unph{\nabla}_b \ngen  - \left( \unit^a \unph{\nabla}_a \unph{\spacemet}^b{}_c + \unph{\spacemet}^a{}_c \K_a{}^b \right) \unph\tgen_b   + \Om \unph{\spacemet}_{bc} \unit^a  \K_a{}^b   \ngen  + v_c & = 0, \label{hypngamdelta2} \\
-2 \unit^b \unph \nabla_b \ngen +  \lap{\unph{R}_{ab}} \unit^a \unph \tgen^b+ u  & = 0,\label{hypnndelta2}
\end{align}
where $\{v_c,u\}$ are given fields determined by the original components $\unit^a \unph \spacemet^b{}_c \delta g_{ab}, \unit^a \unit^b \delta g_{ab}$ respectively. 
Equations \eqref{hypngamdelta2}-\eqref{hypnndelta2} form a symmetric-hyperbolic system of linear PDEs (see subsection \ref{gaugepdehyp} of Appendix \ref{apphyperbolic} for details). This means that it admits a well-posed initial value problem and therefore we can always find a solution for given initial data, which moreover, happen to be unconstrainted (cf. Appendix \ref{apphyperbolic}).

Note that in terms of the fields defined in Section \ref{secregularization}, equation \eqref{eqalinFGgauge} means that there exists a choice of gauge $\{\tau'_{ab},\tau'_a,\tau' \}$ such that $\tau'_a = 0 $ and $\tau'= 0$. Actually the inhomogeneous terms $v_c $ and $ u $ in the system \eqref{hypngamdelta1}-\eqref{hypngamdelta2} coincide, respectively, with terms $\tau_c$ and $\tau$ of the  original field configuration. A similar analysis can be carried out in terms of the fields $\{\tau_{ab},\tau_a,\tau \}$ just by multiplying by the adequate power of an $ \lap$ factor (cf. \eqref{eqtauabdg}-\eqref{eqtaudg}).

\section{Fall-off analysis of background fields in presence of cosmological constant}\label{seceinstein}
In this section, we discuss the asymptotic behaviour of the background gravitational fields. We will use this result and the fall-off behaviour of the linearized field from section \ref{secregularization} to compute symplectic flux in section \ref{secpresymplec}.

Assume that \eqref{unphEFE} holds at a neighbourhood of the boundary $\scri = \{ \Om = 0 \}$. We now examine the fall-off conditions that arise from this assumption, which will later be applied in Section \ref{secpresymplec} to obtain a regular action functional up to the boundary, and consequently, a presymplectic potential. To this end, we consider the $3+1$ decomposition of equation \eqref{unphEFE} as outlined in the introduction, specifically, associated to the $\Omega = \const$ foliation, while using the notation from expressions \eqref{notation1}, \eqref{notation2}, and \eqref{notation3}.

The lowest order term in \eqref{unphEFE} falls-off with $\Om^{-2}$ . Hence regularizing higher order divergent term, we obtain
\begin{equation}\label{eqfofflapse}
\unph{g}^{cd}\unph{\nabla}_c \Om \unph{\nabla}_d \Om  + \el^{-2} = -\lap^{-2} + \el^{-2} 
= \mathcal{O}(\Om).
\end{equation}
We have discussed the regularization of this term by introducing $f$ in \eqref{deff}. Alternatively, we can introduce an auxiliary field $\a$ such that
\begin{equation} \label{28XI24.03}
    \lap^{-1} =: \el^{-1} - \Om \a,
\end{equation}
and  therefore
\begin{equation}\label{eqfofflapse2}
\unph{g}^{cd}\unph{\nabla}_c \Om \unph{\nabla}_d \Om  + \el^{-2} = -\lap^{-2} + \el^{-2} = \Om \a (2 \el^{-1} - \Om \a).
\end{equation}
Note that $\a$ and $\tilde{f}$ are related by
\begin{equation}\label{afrel}
\tilde{f} = 2\el^{-1} \a - \Om \a^2. 
\end{equation}
For convenience, in this section, we will use $\unph a$ instead of $\unph f$.

Now \eqref{unphEFE} reads
\begin{equation}\label{unnphEFE1}
\unph R_{ab} = \Om^{-1} \left\lbrace 
-2 \unit_b \unph{\nabla}_a \lap^{-1}  - \frac{2}{\lap} \K_{ab} - \unph{g}_{ab}  \unit^c \unph \nabla_c \lap^{-1} -\frac{\ug_{ab}}{\lap} \k + 3 \A \ug_{ab}
\right\rbrace,
\end{equation}
where we have used the definition of $\tilde{\nu}_{a}, \tilde{K}_{ab}, \tilde{k}_{ab}$ from Eqs. \eqref{notation1}, \eqref{notation3}.

Next, we write the tangent components of \eqref{unnphEFE1} in terms of the intrinsic geometry of the constant $\Om$ leaves.
Recall that the Gauss identity reads
\begin{equation}\label{gaussid}
{\unph{R}_{abcd}\unph{\gamma}^{a}{}_{p} \unph{\gamma}^{b}{}_{q}\unph{\gamma}^{c}{}_{r}\unph{\gamma}^{d}{}_{s}} = \unph{r}_{pqrs} - \left(\k_{ps} \k_{qr} - \k_{pr} \k_{qs} \right),
\end{equation}
where 
{$\unph{r}_{pqrs}$} is the Riemann tensor of $\uspcmet$. Taking the $\uspcmet$-trace of \eqref{gaussid}, we obtain
\begin{equation}\label{gaussid2}
\tilde{\gamma}^{q}{}_{b} \tilde{\gamma}^{s}{}_{d}\unph{R}_{qs} + \unph{R}_{pbrd} \unit^p \unit^r = \unph r_{bd} - \k^2_{bd} + \k \k_{bd}\,.
\end{equation}
Here $\unph r_{bd}$ is the Ricci tensor of $\uspcmet$, $\k^2_{bd} = \k_{ba}\k^a{}_d$ and $\k = \uspcmet^{ab} \k_{ab}$. The term $\unph{R}_{abcd} \unit^a \unit^c$ can be written in terms of the electric part of the Weyl tensor $\elec_{bd} := \unph{C}_{abcd} \unit^a \unit^c$ as
\begin{equation}\label{eqriemnn}
\unph{R}_{abcd} \unit^a \unit^c = \elec_{bd} - \frac{1}{2} \tilde{\gamma}^{q}{}_{b} \tilde{\gamma}^{s}{}_{d}\unph{R}_{qs} + \frac{1}{2} \uspcmet_{bd} \unph{R}_{ac} \unit^ a \unit^c + \frac{ \unph R}{6} \uspcmet_{bd}.
\end{equation}
Now recall the well-known fact that the rescaled Weyl tensor $\relec_{ab} = \Om^{-1} \elec_{ab}$ is a regular object at~$\scri$. In a nutshell, this can {be understood as} a consequence of the contracted Bianchi identity for the Weyl tensor. In the absence of the source, i.e. $T_{ab}=0$, contracted Bianchi identity reduces to 
\bea 
\nabla_a C^a{}_{bcd} = 0.
\eea
Therefore, from the properties of conformal rescaling, it follows that
\begin{equation}
\unph \nabla_a (\Om^{-1} C^a{}_{bcd}) = \Om^{-1} \nabla_a C^a{}_{bcd} = 0.
\end{equation}
Equation $\unph \nabla_a (\Om^{-1} C^a{}_{bcd}) = 0$ is part of the hyperbolic system of Friedrich's conformal field equations (see the review \cite{Friedrich:2014rpa}) which is smoothly extendable to $\scri$. Thus, all components of $\Om^{-1} C^a{}_{bcd}$ regularly extend to $\scri$ so we can assume regularity of $\relec_{ab} = \Om^{-1} \elec_{ab}$. In terms of this new variable, equation \eqref{eqriemnn} becomes
 \begin{equation}
\unph{R}_{abcd} \unit^a \unit^c = \Om \relec_{bd} - \frac{1}{2} \tilde{\gamma}^{q}{}_{b} \tilde{\gamma}^{s}{}_{d}\unph{R}_{qs} + \frac{1}{2} \uspcmet_{bd} \unph{R}_{ac} \unit^ a \unit^c + \frac{ \unph R}{6} \uspcmet_{bd}.
\end{equation}
Inserting this back into \eqref{gaussid2}, after some algebra, we get the following identity 
\begin{equation}\label{gaussid3}
\tilde{\gamma}^{q}{}_{b} \tilde{\gamma}^{s}{}_{d}\unph R_{qs} = -2 \Om \relec_{bd} - \left( \unph R_{ac}\unit^a\unit^c + \frac{\unph R}{3} \right) \uspcmet_{bd} + 2 \left( \unph r_{bd} - \k^2_{bd} + \k \k_{bd} \right).
\end{equation}
Now taking the normal projection of the conformal Einstein equation \eqref{unnphEFE1}, we get
\begin{align} \label{28XI24.01}
\unph R_{ab}\unit^a \unit^b + \frac{\unph R}{3} & = \Om^{-1} \left\lbrace
\unit^c \unph\nabla_c \lap^{-1} - \frac{\k}{\lap} + \A
\right\rbrace.
\end{align}
Similarly, taking the spatial projection of the conformal Einstein equation \eqref{unnphEFE1}, we have
\begin{align} \label{28XI24.02}
\tilde{\gamma}^{q}{}_{b} \tilde{\gamma}^{s}{}_{d}\unph R_{qs} & = \Om^{-1} \left\lbrace
 -\frac{2}{\lap} \k_{bd} -{\uspcmet_{bd}}  \unit^c \unph \nabla_c \lap^{-1} - \frac{\uspcmet_{bd}}{\lap} \k + 3 \A \uspcmet_{bd}
\right\rbrace.
\end{align}
Putting Eqs. \eqref{28XI24.01}, \eqref{28XI24.02} back into \eqref{gaussid3} yields
\begin{equation}\label{unphEFEtt}
\Om\relec_{bd} - \unph r_{bd} + \k^2_{bd} - \k \k_{bd} = \Om^{-1} \left\lbrace 
\frac{1}{\lap} (\k_{bd} + \k \uspcmet_{bd}) - 2 \A \uspcmet_{bd}
\right\rbrace .
\end{equation}
Thus, using \eqref{28XI24.03}, we obtain
\begin{equation}\label{unphEFEtt1}
\Om \relec_{bd} - \unph r_{bd} + \k^2_{bd} - \k \k_{bd} = \Om^{-1} \el^{-1}\left\lbrace 
 \k_{bd} + \k \uspcmet_{bd} - 4 \a \uspcmet_{bd}
\right\rbrace -
\left\lbrace 
 \a(\k_{bd} + \k \uspcmet_{bd}) - 2 \a^2 \uspcmet_{bd} 
\right\rbrace .
\end{equation}

Let us denote {by the superscript $^{\tf}$ the trace-free} part of $2$-tensors $X$ in spacelike slices,
  \begin{equation}
     X^{\tf}_{ab} := X_{ab} -\frac{1}{3}\unph\gamma^{cd}X_{cd}  \uspcmet_{ab}  .
  \end{equation}
With this notation, the {regularization of} trace-free and trace parts 
of right hand side of the \eqref{unphEFEtt} require the fall-off conditions for 
$k_{ab}$
  \begin{align}
     \k_{bd}^{\tf} &= \Om  \X_{bd},\label{eqfoffktf}\\
       \a  &=  \frac{\k}{3}  + \Om  \b,\label{eqfoffa}
 \end{align}
where $\X_{bd}$ is an auxiliary trace-free tensor and $\b$ an auxiliary scalar, both regular at $\Om = 0$. Then, inserting \eqref{eqfoffktf}-\eqref{eqfoffa} back into \eqref{unphEFEtt1} yields, after canceling terms,

 \begin{equation}\label{unphEFEtt2}
     \Om\relec_{bd} -\unph r_{bd} + \Om^2 \X^2_{bd} = (\el^{-1} - \b \Om^2) \X_{bd} + 2 \b (-2 \el^{-1} + \b \Om^2) \uspcmet_{bd} .
 \end{equation}
From the trace-free and trace parts of \eqref{unphEFEtt2} we obtain, respectively, 

\begin{align}
    \el^{-1}\, \X_{bd}   & =- \unph r_{bd}^{\tf}+ \Om\, \relec_{bd} + \Om^2\left( b \X_{bd} + \X^{2\,\tf}_{bd} \right),\label{eqfoffchi}\\
     4 \el^{-1} \b   & =\frac{\unph r}{3}+ \Om^2\Bigl(2 \b^2 - \frac{1}{3}\, \X^{ac}\X_{ac}\Bigr).\label{eqfoffb}
 \end{align}

In summary, inserting \eqref{eqfoffktf} into \eqref{eqfoffchi}, as well as \eqref{eqfoffb}  into \eqref{eqfoffa} and then the result into \eqref{28XI24.03}, we obtain the following fall-off conditions for the fields
\begin{align}
    \k_{bd}^{\tf} & = -\el \unph r_{bd}^{\,\tf} \,\Om + {\el}\relec_{bd}\,\Om^2 + \mathcal{O}(\Om^3),\label{trfreefalloff}\\
       \lap^{-1} & =\el^{-1} -\frac{\k}{3}\, \Om - \el\frac{\unph r}{12}\,  \Om^2 + \mathcal{O}(\Om^4) .\label{trfalloff}
\end{align}
We remark that it is possible to keep generating further 
terms in \eqref{trfreefalloff}-\eqref{trfalloff} by introducing 
additional auxiliary fields and evaluating the conformal Einstein 
equation. However, for our purposes, it suffices with the orders 
appearing in \eqref{trfreefalloff}-\eqref{trfalloff}. Additionally, 
note that these are not Taylor series expansions of field variables. Each term multiplying a 
power of $\Om$ is not a coefficient of an expansion in $\Om$ since 
they generally also depend on $\Omega$. This formulation is preferable for our purposes, as it allows for 
the exact cancellation  of certain terms that 
arise in the computation of the presymplectic potential. One could 
choose to carry out the same analysis by generating a Taylor series 
expansion. However, it is 
convenient to introduce auxiliary variables in a gauge-free approach, 
because a Taylor series expansion in 
an arbitrary gauge has a rather involved form. Taylor series 
expansion of field variables in the Fefferman-Graham gauge for asymptotically AdS and 
dS spacetimes can be found in \cite{Hollands:2005wt} and 
\cite{PremaBalakrishnan:2019jvz} respectively.\footnote{Note that the `Dirichlet' or `reflective' boundary condition, used the references \cite{Hollands:2005wt,PremaBalakrishnan:2019jvz}, contributes to the vanishing gravitational flux across the boundary. }  

Note that \eqref{trfreefalloff} provides insight into the structure of the asymptotic phase space. 
Specifically, the lowest non-trivial order $\Om$ is determined by the boundary metric, which corresponds to 
the initial configuration state, while $\relec$ appears at the subleading order in the expansion, effectively 
acting as a normal derivative and thus representing a momentum-like quantity. On the other hand, due to the 
gauge nature of $\lap$, the sub-leading terms in \eqref{trfalloff} are gauge-dependent quantities and do not essentially contribute to the dynamical evolution of the system.

\section{Symplectic flux}\label{secpresymplec}
In this section, we calculate the presymplectic potential and presymplectic current leading to the formula of the symplectic flux. We will compute the presymplectic potential from the first principle starting from the action.
We define the following action terms
\begin{equation}
    S_{EH} :=\frac{1}{16\pi G}\int_\man |\phy g|^{1/2} (\phy R - 6 \el^{-2}),\quad s_{GH}:= - \frac{1}{16\pi G}\int_{\partial\man} |\phy \spacemet|^{1/2} 2 \phy k,\quad s_{ct} := \frac{{1}}{16\pi G}\int_{\partial\man} |\phy\spacemet|^{1/2} (4 \el^{-1} - \el \phy r),
\end{equation}
respectively known as Einstein-Hilbert action and Gibbons-Hawking and  counterterms. The total action then reads
\begin{equation}\label{eqtotalaction}
    S = S_{EH} + s_{GH} + s_{ct}=\frac{1}{16\pi G}\bigg(\int_\man |\phy g|^{1/2} (\phy R - 6 \el^{-2}) - \int_{\partial\man} |\phy \spacemet|^{1/2} 2 \phy k  + \int_{\partial\man} |\phy\spacemet|^{1/2} (4 \el^{-1} - \el{\phy r})\bigg).
\end{equation}
 In order to calculate the boundary terms at $\scri$, we shall first consider integration within the domain $\dom_\epsilon$ with boundary $\partial \dom_\epsilon = \{ \Om = \epsilon >0\}$. 
 Later we shall take the limit $\epsilon \rightarrow 0$ so that $\lim_{\epsilon \rightarrow 0} \dom_\epsilon = M$. For simplicity, we will first assume that $\partial \man$, as well as $\partial \dom_\epsilon$, are compact, without a boundary. However, for clarity, we keep track of all the boundary terms. Only at the end of the computation, we will use the absence of the boundary of $\partial \dom_\epsilon$.

  In the Gibbons-Hawking term, the sign depends on the convention chosen for $k$ and the causal character of $\partial \dom_\epsilon$. This, in turn, fixes a unique sign for the counterterm, since this is meant to cancel divergences of the action.  
 In our case this is fixed by $\partial \dom_\epsilon$ being spacelike and $k_{ab} = \spacemet^c{}_a\spacemet^d{}_b \nabla_c n_b$  constructed with $n^a = N g^{ab}\nabla_b \Om$, pointing outward at $\partial \dom_\epsilon$ (see comment below equation \eqref{eqboundEH}).  Our sign convention matches with that of \cite{Compere:2020lrt}, but other conventions may be also used \cite{Anninos:2010zf, Compere:2019bua, Compere:2008us}.
 
 Now consider a linear variation of the metric $g \rightarrow g + \delta g$. We keep $\Om$ as an invariant foliating function, namely $\delta \Om = 0$. Note that under this assumption
\begin{equation}\label{deltaphyunit}
    \delta \phyunit_a = \delta (\phylap \nabla_a \Om) = \frac{\phyunit_a}{\phylap}\delta \phylap.
\end{equation}
From the $3+1$ decomposition, we also note that
\begin{equation}\label{nornordeltagam}
     0 = \delta(\phyunit^a\phyunit^b \spacemet_{ab}) =  \delta \phyunit^a \phyunit^b \spacemet_{ab} + \phyunit^a \delta \phyunit^b \spacemet_{ab} + \phyunit^a \phyunit^b  \delta \spacemet_{ab}= \phyunit^a \phyunit^b  \delta \spacemet_{ab},
\end{equation}
which in turn also implies 
 \begin{equation}
   \spacemet^c{}_a \spacemet^d{}_b \delta g_{cd} =  \spacemet^c{}_a \spacemet^d{}_b (-\delta \phyunit_c \phyunit_d -  \phyunit_c \delta \phyunit_d +  \delta \spacemet_{cd} ) = \spacemet^c{}_a \spacemet^d{}_b  \delta \spacemet_{cd} .
 \end{equation}

We calculate the contribution to the presymplectic potential of $S_{EH}, s_{GH}$ and $s_{ct}$ separately by obtaining variations of their respective Lagrangian densities $\mathcal L_{EH}, \mathcal L_{GH}$ and $\mathcal L_{ct}$. Starting with $\mathcal L_{EH}$
\begin{align}
   16\pi G \delta \mathcal{L}_{EH} & =  |\phy g|^{1/2} \delta \phy R_{ab} \phy g^{ab} + |\phy g|^{1/2} \phy R_{ab} \delta g^{ab} +  \delta |\phy g|^{1/2} (\phy R - 6 \el^{-2}),\\
    & =  |\phy g|^{1/2} \delta \phy R_{ab} \phy g^{ab} + |\phy g|^{1/2} (\phy R_{ab}- \frac{1}{2}\phy R {g_{ab}}+3 \el^{-2} {{g}_{ab}}){\delta g^{ab}} = |\phy g|^{1/2} \delta \phy R_{ab} \phy g^{ab},
\end{align}
where the last equality holds on-shell. Taking into account that
\begin{align}
 \delta \phy{R}_{ab} = -\frac{1}{2} \phy{\nabla}_a \phy{\nabla}_b \delta \phy{g} -\frac{1}{2} \phy{\Box} \delta \phy{g}_{ab} + \phy{\nabla}^c \phy{\nabla}_{(a} \delta \phy{g}_{b)c},
\end{align}
we obtain 
\begin{align}\label{eqboundEH}
    |\phy g|^{1/2} \delta \phy R_{ab} \phy g^{ab}  =  |\phy g|^{1/2} \phy \nabla_a \phy g^{ac}(\phy \nabla^b \delta \phy g_{bc} - \phy g^{de} \phy \nabla_a \delta \phy g_{de}).
\end{align}

Now observe that, in the unphysical picture, the unit vector $\unit^a$ is outward pointing at $\scri$, because $\unit^a\nabla_a\Omega =\tilde  N^{-1} \unit_a \unph{g}^{ab} \unit_b = - \tilde N^{-1} <0$ and $\Omega$ grows inward from $\scri$. Similarly, because $\Omega>0$ in $M$, the physical unit vector $\phyunit^a = \Omega \unit^a$ is the outward oriented at $\partial \dom_\epsilon$. Then, by the Gauss theorem (cf. Appendix \ref{appstokes}),
\begin{equation}\label{EHboundtermderivation}
  \int_{\dom_\epsilon} \!\!\!|\phy g|^{1/2} \phy \nabla_a \phy g^{ac}(\phy \nabla^b \delta \phy g_{bc} - \phy g^{de} \phy \nabla_c \delta \phy g_{de})  
  =  \int_{\partial\dom_\epsilon}\!\!\!\!\!\!|\phy \spacemet|^{1/2} (-\phyunit^c) (\phy \nabla^b \delta \phy g_{bc} - \phy g^{de} \phy \nabla_c \delta \phy g_{de}) 
  = 16 \pi G\int_{\partial\dom_\epsilon} \!\!\!\!\!\!\theta_{EH},
\end{equation}
where $\theta_{EH}$ denotes the contribution of the Einstein-Hilbert term to the presymplectic potential
\begin{align}\label{thetaEH1}
    16 \pi G \theta_{EH} & = - |\phy \spacemet|^{1/2} \phyunit^a (\phy \nabla^b \delta \phy g_{ab} - \phy g^{cd} \phy \nabla_a \delta \phy g_{cd}).
\end{align}
Note that, following notation \eqref{notation3},  $\phy k = \spacemet^{ab}\phy k_{ab} = \phy g^{ab}\phyK_{ab} = \phyK$, where recall that both $k_{ab}$ and $K_{ab}$ are defined using  the outward-oriented normal. Then we make the following observation 
\begin{align}
    \delta \phyk = \delta \phyK = \delta \phy g^{ab} \phyK_{ab} + \phy g^{ab} \delta \phyK_{ab} =  -\delta \phy g_{ab} \phyK^{ab} + \phy g^{ab} \delta \phyK_{ab} ,
\end{align}
where we can write
\begin{align}
    \phy g^{ab} \delta \phyK_{ab} & = \phy g^{ab} \left(\phy \nabla_a \delta \phyunit_b - \frac{\phyunit^c}{2}(\phy\nabla_a \delta \phy g_{bc} + \phy\nabla_b \delta \phy g_{ac} -\phy\nabla_c \delta \phy g_{ab})\right),\\ 
    & = \phy g^{ab} \phy \nabla_a \delta \phyunit_b - \frac{1}{2}\phy g^{ab}\phy\nabla_a (\phyunit^c \delta \phy g_{bc}) +\frac{1}{2}\phy g^{ab}\delta \phy g_{bc}\phy\nabla_a \phyunit^c  -\frac{\phyunit^c}{2}\phy g^{ab} (\phy\nabla_b \delta \phy g_{ac} - \phy\nabla_c \delta \phy g_{ab}),\\
    & = \phy g^{ab} \phy \nabla_a \delta \phyunit_b - \frac{1}{2}\phy g^{ab}\phy\nabla_a (\phyunit^c \delta \phy g_{bc}) + \frac{1}{2}\phyK^{ac}\delta \phy g_{ac}  -\frac{\phyunit^c}{2}\phy g^{ab} (\phy\nabla_b \delta \phy g_{ac} - \phy\nabla_c \delta \phy g_{ab}),
\end{align}
and hence
\begin{align}
     \delta \phyk =  -\frac{1}{2}\delta \phy g_{ab} \phyK^{ab}  + \phy g^{ab} \phy \nabla_a \delta \phyunit_b - \frac{1}{2}\phy g^{ab}\phy\nabla_a (\phyunit^c \delta \phy g_{bc}) -\frac{\phyunit^c}{2}\phy g^{ab} (\phy\nabla_b \delta \phy g_{ac} - \phy\nabla_c \delta \phy g_{ab}).
\end{align}
Putting this back into \eqref{thetaEH1} yields
\begin{align}
       {16\pi G}\theta_{EH} =  |\spacemet|^{1/2}\left(2 \delta \phyk + \delta \phy g_{ab} \phyK^{ab} - 2 \phy g^{ab} \phy \nabla_a \delta \phyunit_b  + \phy g^{ab}\phy\nabla_a (\phyunit^c \delta \phy g_{bc})\right),
\end{align}
which after noticing
\begin{align}
    \phyunit^c \delta \phy g_{bc} = \delta \phyunit_b - \phyunit_b \phyunit^c \delta \phyunit_c + \phyunit^c \delta \spacemet_{bc} =  2 \delta \phyunit_b+ \phyunit^c \delta \spacemet_{bc},
\end{align}
we obtain
\begin{align}
       {16 \pi G}\theta_{EH} =  |\spacemet|^{1/2}\left(2 \delta \phyk + \delta \phy g_{ab} \phyK^{ab}   + \phy g^{ab}\phy\nabla_a (\phyunit^c \delta  \spacemet_{bc})\right).
\end{align}
Now, taking into account  $K^{ab} \phyunit_b = 0$ and \eqref{deltaphyunit}, we obtain,
\begin{equation} \label{20II25.01}
    \phyK^{ab} \delta \phy g_{ab} = \phyK^{ab}(-\delta \phyunit_a \phyunit_b- \phyunit_a \delta \phyunit_b + \delta \spacemet_{ab}) = \phyK^{ab} \delta \spacemet_{ab} = \phyk^{ab}\delta \spacemet_{ab} - \phyunit_c \phyK^{cb} \phyunit^a \delta \spacemet_{ab}.
\end{equation}
From \eqref{nornordeltagam}, we also have
\begin{align} \label{20II25.02}
     \phyunit_c \phyK^{cb} \phyunit^a \delta \spacemet_{ab} = \phyunit_c (\phy \nabla^c \phyunit^b) \phyunit^a \delta \spacemet_{ab} = - \phyunit^c \phyunit^b (\phy \nabla_c \phyunit^a \delta \spacemet_{ab}).
\end{align}
Using \eqref{20II25.01}, \eqref{20II25.02}, we finally find 
\begin{align}
16 \pi G \theta_{EH} =  |\spacemet|^{1/2}\left(2 \delta \phyk + \phyk^{ab}  \delta \phy \spacemet_{ab}   + \spacemet^{ab}\phy \nabla_a (\phyunit^c \delta  \spacemet_{bc})\right).\label{thetaEH2}
\end{align}
Note that $\phyunit^c \delta  \spacemet_{bc}$ is a tangent vector of $\partial \dom_\epsilon$ (cf. \eqref{nornordeltagam}), thus $ \spacemet^{ab}\phy \nabla_a (\phyunit^c \delta  \spacemet_{bc}) = \spacemet^{ab}\phy D_a (\phyunit^c \delta  \spacemet_{bc})$
is a total derivative of $\partial \dom_\epsilon$ and therefore it will not contribute to the integral (as long is $\partial \dom_\epsilon$ compact.)  We however keep track of this term by defining $u^a := \gamma^{ab} \delta  \spacemet_{bc} \phyunit^c$.
Hence
    \begin{align}\label{thetaEH3}
16 \pi G \theta_{EH} =  |\spacemet|^{1/2}\left(2 \delta \phyk + \phyk^{ab}  \delta \phy \spacemet_{ab}   + D_a u^a\right).
\end{align}
The contribution of the Gibbons-Hawking and counter-terms is straightforwardly computable, since they both are boundary integrals
\begin{align}
   16 \pi  G\theta_{GH} & = -2  |\spacemet|^{1/2}\delta k - |\spacemet|^{1/2}k \spacemet^{ab}\delta\spacemet_{ab},\label{thetaGH1}\\
      16 \pi  G\theta_{ct} & = 2 \el^{-1} |\spacemet|^{1/2} \spacemet^{ab}  \delta \spacemet_{ab} - \el \delta (|\spacemet|^{1/2} \phy r).\label{thetact1}
\end{align}
Combining \eqref{thetaEH2}, \eqref{thetaGH1} and \eqref{thetact1} gives the following expression for the total presymplectic potential 
\begin{align}
    16 \pi G\theta & =16 \pi G( \theta_{EH} +  \theta_{GH} +  \theta_{ct}) \\ & =  |\spacemet|^{1/2} (
 \phyk^{ab}   - k \spacemet^{ab})\delta\spacemet_{ab} 
  + 2 \el^{-1} |\spacemet|^{1/2} \spacemet^{ab}  \delta \spacemet_{ab} - \el \delta (|\spacemet|^{1/2} \phy r)   
 + |\gamma|^{1/2} D_a u^a\label{presympphys}.
\end{align}
Note that the exact cancelation between $\delta k$ term from $\theta_{EH}$, and $\theta_{GH}$ guarantees a well-defined variation principle \cite {PhysRevD.15.2752, Balasubramanian:1999re} and it is reason for introducing the Gibbons-Hawking term into the action.

At this point, we can identify the de Sitter holographic stress tensor \cite{Anninos:2010zf, Balasubramanian:2001nb} analogical to that widely discussed in the anti-de Sitter case \cite{Balasubramanian:1999re}. Neglecting the total derivative term $D_a u^a$ and defining $T^{ab}:= \frac{2}{|\gamma|^{1/2}} \frac{\delta S}{\delta \gamma_{ab}}$, from \eqref{presympphys} we have
\bea 
T^{ab}= \frac{1}{{8\pi G}}\bigg( k^{ab}-k \gamma^{ab} 
 + 2\el^{-1}\gamma^{ab}+\el(r^{ab}-\frac{1}{2}r \gamma^{ab})\bigg).
\eea
 This is also reminiscent of analogous boundary stress tensor in anti-de~Sitter case (see eq. (10) of \cite{Balasubramanian:1999re} for $\mbox{AdS}_{4}$.)

It will be useful to write \eqref{presympphys} in terms of infinitesimal variations of the contravariant metric $(\gamma^{-1})^{ab} = \gamma^{ab}$. Abusing the notation we shall denote 
\begin{equation}
    \delta (\gamma^{-1})^{ab} =   \delta \gamma^{ab},\qquad  \delta (g^{-1})^{ab} =   \delta g^{ab},
\end{equation}
Beware, that in this convention $\delta \gamma^{ab}=-\gamma^{ak}\gamma^{bl}\delta\gamma_{kl}$ and similarly for $g^{ab}$. Thus, \eqref{presympphys} can be easily expressed in terms of $\delta \gamma^{ab}$
\begin{align}
    16 \pi G \theta & = -  |\spacemet|^{1/2} (
 \phyk_{ab}   - k \spacemet_{ab})\delta\spacemet^{ab} - 2 \el^{-1} |\spacemet|^{1/2} \spacemet_{ab}  \delta \spacemet^{ab} - \el \delta (|\spacemet|^{1/2} \phy r) + |\gamma|^{1/2} D_a u^a \label{presympphys2}.
\end{align}

We now transform the above expression to unphysical variables, i.e. those corresponding to $\unph g_{ab} = \Om^2 \phy g_{ab}$ and $\unph \gamma_{ab} = \Om^2 \phy \gamma_{ab}$. Recall that the change of connection is given by the following tensor 
\begin{equation}
    \nabla_a v_b - \unph\nabla_a v_b =  - S_{ab}^c v_c,\qquad S_{ab}^c = -\frac{1}{\Om} ( \unph \nu_b \delta^c_a + \unph \nu_a \delta^c_b - \unph \nu^c \unph  g_{ab} )
\end{equation}
and the physical and unphysical unit normals $n_a$ and $\unph n_a$ satisfy 
\begin{equation*}
    n_a = \Om^{-1} \unph n_a .
\end{equation*}
Then we have the following relation between extrinsic curvatures
\begin{equation}
\phyk_{ab} = \spacemet^c{}_a \spacemet^d{}_b \nabla_a \phyunit_b =  \frac{\k_{ab}}{\Om} + \frac{1}{\Om^2}\frac{\uspcmet_{ab}}{\lap} \Longrightarrow \phyk = \spacemet^{ab} \phyk_{ab} = \Om \k + \frac{3}{\lap}.\label{eqkphyumph}
\end{equation}
To relate the mean intrinsic curvatures, observe that the Ricci tensors satisfy $\phy r_{ab} = \unph r_{ab}$. This is because they are respectively constructed out of the metrics $\gamma_{ab}$ and $\unph \gamma_{ab} = \Om^2 \gamma_{ab}$. Since they are both instrinsic to hypersurfaces with $\Om = \const $, we can consider $\Om$ as a constant for the calculation of $r_{ab}$ and $ \unph r_{ab}$. Then
\begin{equation*}
    \phy r = \spacemet^{ab} \phy r_{ab} = \Om^{2} \uspcmet^{ab} \unph r_{ab} = \Om^2 \unph r,
\end{equation*}
and therefore
\begin{align}
    \delta (|\gamma|^{1/2}r) = \frac{1}{\Om} \delta (|\unph \gamma|^{1/2} \unph r) = \frac{ |\unph \gamma|^{1/2}}{\Om} ( \unph r_{ab} - \frac{1}{2} \unph r \gamma_{ab}  ) \delta \unph \gamma^{ab} + \frac{ |\unph \gamma|^{1/2}}{\Om}  \unph D_a \unph v^a \label{delgamr}
\end{align}
where 
\begin{equation*}
   \unph v^a := \uspcmet^{ad}\uspcmet^{bc}(\unph\D_b \delta \uspcmet_{cd} - \unph \D_d \delta \uspcmet_{bc} ),
\end{equation*}
and the second equality in \eqref{delgamr} is a standard result (cf.\ \cite{Wald:1984rg}), 
parallel to the variation of the spacetime curvature performed above.

Within the integral, the term  $|\gamma|^{1/2} D_a u^a$ is independent from the connection employed (see Appendix \ref{appstokes}, equation \eqref {eqderdensity}), so defining
\begin{equation*}
    \unph u_a  = \Om^{-1} u_a \;,  \quad\unph u^a  = \Om u^a,
\end{equation*}
we may substitute 
\begin{equation*}
    D_a |\gamma|^{1/2}u^a \quad \rightarrow \quad \unph D_a |\unph \gamma|^{1/2}u^a =\frac{|\unph \gamma|^{1/2}}{\Om} \unph D_a \unph u^a. 
\end{equation*}
With these definitions, inserting \eqref{eqkphyumph} and \eqref{delgamr} into  \eqref{presympphys2}, we now have
\begin{align}
    16 \pi G\theta & = 2 \frac{|\uspcmet|^{1/2}}{\Om^3} \left( -\el^{-1} + \lap^{-1} \right) \uspcmet_{ab} \delta \uspcmet^{ab}
     -\frac{|\uspcmet|^{1/2}}{\Om^2}\left( \k_{ab} - \k \uspcmet_{ab} \right) \delta \uspcmet^{ab} \nonumber \\
  &  -\frac{|\uspcmet|^{1/2}}{\Om} \el\left( \unph r_{ab} -\frac{\unph r}{2} \uspcmet_{ab} \right) \delta \uspcmet^{ab}   + \frac{|\unph \gamma|^{1/2}}{\Om} \unph D_a \unph u^a 
     - \frac{1}{\Om} |\uspcmet|^{1/2} \el \unph\D_a \unph v^a. \label{presympec0}
\end{align}

We can gather the total derivative terms, which recall do not contribute to the integral, by defining 
\begin{equation*}
   \unph \beta^a := \unph u^a - \el \unph v^a.
\end{equation*}
Then 
\begin{align}
    16 \pi G \theta & = 2 \frac{|\uspcmet|^{1/2}}{\Om^3} \left( -\el^{-1} + \lap^{-1} \right) \uspcmet_{ab} \delta \uspcmet^{ab}
     -\frac{|\uspcmet|^{1/2}}{\Om^2}\left( \k_{ab} - \k \uspcmet_{ab} \right) \delta \uspcmet^{ab} \nonumber \\
  &  -\frac{|\uspcmet|^{1/2}}{\Om} \el\left( \unph r_{ab} -\frac{\unph r}{2} \uspcmet_{ab} \right) \delta \uspcmet^{ab}   + \frac{|\unph \gamma|^{1/2}}{\Om} \unph D_a \unph \beta^a. \label{presympec1}
\end{align}
We now introduce \eqref{trfalloff} into \eqref{presympec1} and find
\begin{align*}
    16 \pi G\theta & = 
     -\frac{|\uspcmet|^{1/2}}{\Om^2}\left( \k_{ab} - \frac{\k}{3} \uspcmet_{ab} \right) \delta \uspcmet^{ab}  
     -\frac{|\uspcmet|^{1/2}}{\Om} \el\left( \unph r_{ab} -\frac{\unph r}{3} \uspcmet_{ab} \right) \delta \uspcmet^{ab}    
     + \frac{1}{\Om} |\uspcmet|^{1/2} \unph\D_a \unph\beta^a  + \mathcal{O}(\Om)
     \\ & = 
     -\frac{|\uspcmet|^{1/2}}{\Om^2}\left( \k_{ab} - \frac{\k}{3} \uspcmet_{ab} \right) \delta \uspcmet^{ab}  
     -\frac{|\uspcmet|^{1/2}}{\Om} \el\left( \unph r_{ab} -\frac{\unph r}{3} \uspcmet_{ab} \right) \delta \uspcmet^{ab}    
     + \frac{1}{\Om} |\uspcmet|^{1/2} \unph\D_a \unph\beta^a  + \mathcal{O}(\Om)
     \\ & = 
      -\frac{|\uspcmet|^{1/2}}{\Om^2}\,\k^{\tf}_{ab}\, \delta \uspcmet^{ab}  -\frac{|\uspcmet|^{1/2}}{\Om} \el\,\unph r^{\,\tf}_{ab}\, \delta \uspcmet^{ab}    
     + \frac{1}{\Om} |\uspcmet|^{1/2}  \unph\D_a\unph \beta^a + \mathcal{O}(\Om).
\end{align*}
Introducing now \eqref{trfreefalloff}, we are left with
\begin{equation} \label{1XI24.05}
     16 \pi G \theta =   -|\uspcmet|^{1/2} \el \relec_{ab} \delta \uspcmet^{ab}  + \frac{1}{\Om} |\uspcmet|^{1/2} \unph\D_a \unph \beta^a + \mathcal{O}(\Om).
\end{equation}
In summary, we have shown that, on shell, 
   \begin{align}
       \delta S  = \int_{\partial \dom_\epsilon} \theta =   -\frac{\el}{16 \pi G}\int_{\partial \dom_\epsilon} |\uspcmet|^{1/2} \el \relec_{ab} \delta \uspcmet^{ab}  + \frac{\el}{16 \pi G}\frac{1}{\Om} \int_{\partial \dom_\epsilon} |\uspcmet|^{1/2} \unph\D_a \unph \beta^a + \mathcal{O}(\Om).\label{eqdeltaS1}
   \end{align}
   
The boundary $\partial\dom_\epsilon$ splits into two disconnected parts, $\partial\dom_\epsilon=\partial\dom_\epsilon^+\cup\partial\dom_\epsilon^-$, one representing the past boundary and the other the future one. The symplectic potential and the symplectic form is obtained by integrating over one of them, say the future one $\Sigma^+=\partial\dom_\epsilon^+$, and we are especially interested in the limit $\Om\to 0$ when $\Sigma^+\to\scri^+$.

Typically, $\Sigma^+$ is compact without boundary (diffeomorphic to $S^3$ in the de~Sitter-like context). Under this assumption, the total derivative term $\unph\D_a \unph \beta^a$ in \eqref{eqdeltaS1}  does not contribute to the symplectic potential,  
     \begin{align}
       \Theta[\Sigma^+]= \int_{\Sigma^+} \theta
       =-\frac{\el}{16 \pi G}\int_{\Sigma^+} |\uspcmet|^{1/2} \el \relec_{ab} \delta \uspcmet^{ab}  + \mathcal{O}(\Om),\label{eqdeltaS1p}
   \end{align}
   so we can take the limit $\Om \rightarrow 0$
    \begin{align}
       \Theta[\scri^+] 
       = -\frac{\el}{16 \pi G}\int_{\scri^+}|\uspcmet|^{1/2} \relec_{ab} \delta \uspcmet^{ab}.\label{eqdeltaS2}
   \end{align}

One might wish to study behaviour of the symplectic potential $\Theta[\Sigma]$ on an arbitrary subdomain $\Sigma$ of the future boundary $\partial \dom_\epsilon^+$.\footnote{{This may be useful when approaching the future de~Sitter infinity using spatially flat or hyperbolic cosmological models.}} Observe, however, that allowing non-empty $\partial\Sigma$ is not straightforward. One would have to study the fall-off behaviour of the boundary terms $\frac1\Om\int_{\partial\Sigma} \unph\beta_a e^a |\unph h|^{1/2}$ arising from  $\unph \D_a \unph \beta^a$, which leads to a non-trivial discussion going beyond the scope of this paper. This is an interesting question that we shall address elsewhere.

We emphasize that the presence of the counterterms in the action and the fall-off conditions of the field in the conformal Einstein equation play an important role to make the presymplectic potential finite at the boundary. Additionally, note that this result matches standard calculations available in the literature of the symplectic potential that are carried out in the Fefferman Graham gauge. 

{Finally, we compute the presymplectic current, given by,
\begin{eqnarray}
\omega({\delta_{1}\tilde{\gamma}, \delta_{2}\tilde{\gamma}} )=\delta_{1}{\theta}(\delta_{2}\tilde{\gamma})-\delta_{2}{\theta}(\delta_{2}\tilde{\gamma}).
\end{eqnarray}
Assuming the variations $\delta_{1}, \delta_{2}$ commute with each other, we have
\bea  \label{1XI1.03}
\omega({\delta_{1}\tilde{\gamma}, \delta_{2}\tilde{\gamma}} )=\frac{\el}{16\pi G}  \bigg(\delta_{1} (|\unph\gamma|^{1/2}\tilde{e}^{\,ab} )\delta_{2}\tilde{\gamma}_{ab}-\delta_{2} (|\unph\gamma|^{1/2}\tilde{e}^{\,ab} )\delta_{1}\tilde{\gamma}_{ab}\bigg).
\eea
}
Note that from the regularization of the linearized conformal Einstein equation, we have $\delta \tilde{g}_{ab}=\delta \tilde{\gamma}_{ab}=\mathcal{O}(1)$. Therefore, the presymplectic current is non-vanishing on the boundary. Physically, this is attributed to the non-zero gravitational radiation on the boundary $\scri^{+}$ \cite{Wald:1999wa, Poole:2021avh}. It is also well known that in leaky boundary condition, namely where the variation of the field does not vanish, the energy flux is given by the presymplectic potential on the $\scri^{+}$.

In the context of linearized theory, presymplectic potential represents flux formula for isometries of de Sitter background. Note that background Kiling vectors become conformal Killing vector of the background metric on $\scri^{+}$, and $\xi$ is the conformal Killing vector of the de Sitter background boundary. The expression of the flux in learized case becomes
\bea \label{4X124.02}
\bm F_{\xi}= \int_{\scri^{+}} \theta (\delta\tilde{\gamma}_{ab}, \mathcal{L}_{\tilde\xi} \delta\tilde{\gamma}_{ab})= \frac{\el}{16\pi G}\int_{\scri^{+}} |\uspcmet|^{1/2} \delta \relec^{ab} \mathcal{L}_{\tilde\xi}\delta \uspcmet_{ab} ,
\eea
where $\delta \relec^{ab}$ linearized rescaled electric part of the Weyl tensor. This formula has been employed in several literature \cite{Kolanowski:2020wfg, Ashtekar:2015lxa, Bonga:2023eml, Harsh:2024kcl} to compute energy flux of linearized fields around de Sitter background. We note that under the conformal transformation $\tilde{\gamma}_{ab}=\Omega^{2}\gamma_{ab}$, $\tilde{e}^{ab}\to \Omega^{-5}e^{ab}, \tilde{\xi}^{a} \to \xi^{a}$, therefore, the flux formula in equation \eqref{4X124.02} is conformally invariant, i.e. we can express the flux formula in terms of physical variables. It is well-known that linearized field is defined upto a gauge transformation, $\delta g_{ab}\to \delta g_{ab}+\mathcal{L}_{\xi} \bar{g}_{ab}$, where $\bar{g}_{ab}$, in our case de Sitter background metric. One can show that linearized Weyl tensor is gauge invariant on de Sitter background \cite{Stewart:1974uz, Bieri:2013ada}. {Since the flux expression is conformally invariant and linearized Weyl tensor is gauge invariant on background de Sitter space-time, the flux is therefore insensitive to linearized diffeomorphism. This confirms the gauge invariant nature of the flux formula.}

\section{Conclusion}
We analyze conformal Einstein equations to extract fall-off 
conditions of the gravitational fields. These fall-off condition 
are consistent to reproduce gravitational flux at the future 
infinity of de Sitter. We obtain presymplectic structure at the 
boundary in terms of rescaled electric part of the Weyl tensor. 
Though this result is well-known in the context of Fefferman-Graham 
gauge in de Sitter, our approach is unique, in particular, we did 
not impose any gauge condition to extract the presymplectic current 
at $\scri^{+}$. This highlights the gauge covariant nature of the quantity,  enhancing its relevance as a candidate for describing gravitational energy flux. {Our result also emphasizes that a gauge covariant conformal extension of holographic stress tensor on the compact boundary naturally selects electric part of the Weyl tensor in the presymplectic potential computation.}

We have found important differences in regularization of the  
linearized conformal Einstein equation in comparison with 
asymptotically flat case. The presence of cosmological constant introduces a 
higher order divergence term in the linearized equation. The leading order term in the metric perturbation 
does not vanish at $\scri^{+}$, rather the non vanishing perturbation produces non-zero symplectic flux.  
The non-vanishing nature of the perturbation at $\scri^{+}$ is also qualitatively consistent with the 
analysis of gravitational waves in Bondi gauge \cite{Poole:2018koa, Compere:2019bua, Chrusciel:2020rlz, Chrusciel:2021ttc} or generalized harmonic gauge \cite{Ashtekar:2015lxa, Date:2015kma, Compere:2023ktn}. {Our result is also consistent with the linear displacement memory effect in de Sitter space-times. Note that the displacement memory in de Sitter space-times depends on the difference between the final and initial non-vanishing fields at the boundary \cite{Compere:2023ktn}. It will also be interesting to explore the gravitational memory effect in terms of the perturbed electric part of the Weyl tensor \cite{Bieri:2015jwa} with the choice of our `minimal' fall-off behaviour of the linearized fields.} To us, the 
constraint in equation \eqref{25XI24.01} is interesting and non-trivial 
for non-zero cosmological constant setting. Our results do not 
rigorously show that the field variables we have introduced in the 
linearized conformal Einstein equation form a hyperbolic PDE 
system. We wish to return to this problem in future.

\section*{Acknowledgements}

We gratefully thank David Hilditch and Tomáš Ledvinka for useful discussions. The work of JH is supported in part by MSCA Fellowships CZ - UK2 $(\mbox{reg. n. CZ}.02.01.01/00/22\_010/0008115)$
from the Programme Johannes Amos Comenius co-funded by the European Union. CPN acknowledges financial support from Departamento de Matemática Aplicada a las TIC - D540. Authors acknowledge the support from Czech Science Foundation Grant 22-14791S.
\appendix

\section{Well-posedness of symmetric-hyperbolic quasilinear PDEs}\label{apphyperbolic}

In this appendix we summarize existence, uniqueness and stability results for quasilinear, symmetric-hyperbolic systems of PDEs. This is based on the results in Chapter 12 of \cite{Kroon:2016ink}, in turn based on Kato \cite{Kato1975TheCP} and Friedrich's \cite{Friedrich:1986qfi} theorems for existence of solutions of quasilinear hyperbolic systems.

Consider a domain $\mathcal{D}$ with topology $\mathbb{R} \times \Sigma$ and coordinates  $\{ \Om,x^i \}$ adapted to a foliation with $\mathcal{D}_\Om := \{\Om = \const.\}$ leaves, each one of which is homemorphic to $\Sigma$. 
Let $\mathbf{U}: \mathcal{D} \longrightarrow \mathbb{R}^n $ be an $\mathbb{R}^n$-valued field  and for each fixed value of $\Om$ define $ \mathbf{u}_\Om:= \mathbf{U}(\Om,\cdot)$ as a field $\mathbf{u}_\Om: \mathcal{D}_\Om \longrightarrow \mathbb{R}^n$. We introduce the Sobolev norm 
\begin{equation}
    || \mathbf{u}_\Om ||_{\mathcal{D}_\Om,m} = \left( \sum_{k=0}^m \sum_{\alpha_1,\cdots,\alpha_k = 0}^3 \int_{\mathcal{D}_\Om}|\partial_{x^{\alpha_k}}\cdots \partial_{x^{\alpha_1}}\mathbf{u}_\Om|^2 \right)^{1/2}
\end{equation}
where $|\mathbf{u}_\Om|^2 = \left\langle \mathbf{u}_\Om,\mathbf{u}_\Om\right\rangle$ stands for the usual $\mathbb{R}^n$ norm. The  Sobolev space $H^m(\mathcal{D}_ \Om,\mathbb{R}^n)$ is the Banach space of $\mathbb R^n$-valued fields of $\mathcal{D}_\Om$ with finite norm $ || \cdot ||_{\mathcal{D}_\Om,m}$ completed with the limit points of its Cauchy sequences. Recall that it is well-known (see e.g. \cite{taylor1996}) that $H^m(\mathcal{D}_ \Om,\mathbb{R}^n) \subset C^{m-2}(\mathcal{D}_ \Om,\mathbb{R}^n)$.

Now consider the quasilinear system of PDEs 
\begin{equation}\label{PDE}
    A^\Om(\Om,x;\mathbf{U})\partial_\Om \mathbf{U} + A^i(\Om,x;\mathbf{U}) \partial_i \mathbf{U} + B(\Om,x;\mathbf{U})= 0,
\end{equation}
where $ A^\Om(\Om,x;\mathbf{U})$, $A^i(\Om,x;\mathbf{U})$  are matrices and $B(\Om,x;\mathbf{U})$ is a vector. Their entries may depend nonlinearly on the coordinates and the field $\mathbf{U}$, but not on its derivatives. We assume that \eqref{PDE} is symmetric, i.e. $ A$, $A^i$ are symmetric matrices, and hyperbolic, i.e. that we can find scalars $(\sigma_\Om, \sigma_i)$ such that the combination $A^\Om \sigma_\Om + A^i \sigma_i$ is a positive definite matrix. We aim to establish existence of the Cauchy problem of \eqref{PDE} with initial data
\begin{equation}\label{cauchydata}
    \mathbf{u}_{\Om_0}= \mathbf{u}_0(x^i) \in H^m(D_{\Om_0},\mathbb{R}^n)
\end{equation}
at some initial surface $D_{\Om_0} = \{\Om = \Om_0 \}$. We assume that $\det A^\Om(\Om_0,x;\mathbf u_{\Om_0}) \neq 0$, so that all transversal derivatives $\partial_\Om \mathbf{U}\mid_{\Om_0}$ can be obtained from the initial data via \eqref{PDE}. This implies that no constraint equations need to be imposed on the initial data. This  assumption can be removed if suitable constraints are introduced.

The existence and uniqueness result is as follows. Consider the Cauchy problem \eqref{PDE}-\eqref{cauchydata} with $m \geq 4$. Assume that $A^\Om(\Om_0,x;\mathbf u_{\Om_0})$ is bounded away from zero for some $\delta >0$, namely, there exists a $\delta >0$ such that $\left\langle \mathbf{Z}, A^\Om  (\Om_0,x;\mathbf u_{\Om_0}) \mathbf{Z} \right\rangle >\delta \left\langle \mathbf{Z},  \mathbf{Z} \right\rangle$ for all $\mathbf{Z} \in \mathbb{R}^n$. Then there exists some $\Om_1 \in \mathbb{R}$ and unique solution $\mathbf{U}(\Om,x^i)$ to the Cauchy problem  \eqref{PDE}-\eqref{cauchydata} such that $\mathbf{U} \in C^{m-2}([\Om_0,\Om_1]\times \Sigma,\mathbb{R}^n)$. Moreover, $A^\Om(\Om,x^i,\mathbf{U})$ remains bounded away from zero in $[\Om_0,\Om_1]\times \Sigma$.

To state the stability results, denote $B_r(\mathbf{u}_{\Om_0})$ to the ball of  $H^m(\mathcal{D}_{\Om_0},\mathbb{R}^n) $ centered at $\mathbf{u}_{\Om_0}$ with radius $r$ wrt to the norm $||\cdot||_{\mathcal{D}_{\Om_0},m}$, intersected with the subspace of functions $\mathbf{u}_{\Om_0} \in H^m(\mathcal{D}_{\Om_0},\mathbb{R}^n)$ for which $A^\Om  (\Om_0,x;\mathbf u_{\Om_0})$ is bounded away from zero by some $\delta>0$.
Then 
\begin{enumerate}
    \item There exists some $r>0$ such that $\Om_1\in \mathbb{R}$ can be chosen so that all solutions with initial conditions in $B_r(\mathbf{u}_{\Om_0})$ exist for the same ``time" interval  $[\Om_0,\Om_1] \subset \mathbb{R}$.
    \item For any Cauchy sequence $\{\mathbf{u}_{\Om_0}^n \} \subset B_r(\mathbf{u}_{\Om_0})$ converging to $\mathbf{u}_{\Om_0}$ on $D_{\Om_0}$, the corresponding solutions $\mathbf{U}^n$ satisfy that $\mathbf{u}^n_\Om =\mathbf{U}^n(\Om,\cdot)$ converge uniformly to  $\mathbf{u}_\Om =\mathbf{U}(\Om,\cdot)$ on each Cauchy slice $\mathcal{D}_\Om$ for all $\Om \in [\Om_0,\Om_1]$.
    \item Given a solution $\mathbf U$ that exists for some interval  $[\Om_0,\Om_1] \subset \mathbb{R}$, then for $r>0$ sufficiently small, all solutions with initial data on $B_r(\mathbf{u}_{\Om_0})$ exist on $[\Om_0,\Om_1]$.
\end{enumerate}



\subsection{Hyperbolicity of \eqref{hypngamdelta2}-\eqref{hypnndelta2}}\label{gaugepdehyp}

We can now apply the existence and uniqueness results stated above  to the PDE system \eqref{hypngamdelta2}-\eqref{hypnndelta2}.  First, we note that a particular type of quasilinear PDE \eqref{PDE} is the linear case, namely,
\begin{equation}
    A^\Om(\Om,x)\partial_\Om \mathbf{U} + A^i(\Om,x) \partial_i \mathbf{U} + B(\Om,x) \mathbf{U} + C(\Om,x)= 0,\label{linPDE}
\end{equation}
where $ A^\Om(\Om,x)$, $A^i(\Om,x)$  and $B(\Om,x)$ are matrices and $C(\Om,x)$ is a vector. Their entries in this case do not depend on $\mathbf{U}$, but they may depend smoothly on the coordinates. 
Recall that we have selected our folitation to have zero shift vector, so that in the adapted coordinates $\{\Om,x^i \}$ we have  $\unit^a = - \lap^{-1}(\partial_\Om)^a$ and $\unph\spacemet^a{}_b = \delta^i{}_j (\partial_{x^i})^a (d x^j)_b $. This simplifies the analysis because it makes the fields $\tgen^i$ to be a linear combination of the fields $\tgen_i$, not involving $\ngen$. 
Comparing \eqref{linPDE} and \eqref{hypngamdelta2}-\eqref{hypnndelta2}, we find that the latter is a PDE system like the former for the field $\mathbf{U} = (\ngen,\unph \tgen^i)$ and the matrices
\begin{equation}
    A^\Om = \lap^{-1}\mathrm{diag}(1,1,1,1),\qquad A^i = \mathrm{diag}(\Om,0,0,0),
\end{equation}
where a global sign has been ommited  and the matrix $B$ and vector $C$ are not specified because neither of them affect the hyperbolicity properties. The system is symmetric because $(A^\Om,A^i)$ are symmetric matrices. Moreover, it is  hyperbolic because for sufficiently small $\Om$ we can find scalars $(\sigma_\Om, \sigma_i)$ such that the combination $A^\Om \sigma_\Om + A^i \sigma_i$ is a positive definite matrix.

 Now consider a prescribed initial field configuration $\mathbf{u}_{\Om_0}$ at some $\{ \Om = \Om_0= \const\}$ initial slice. There is obviously not constraint equations on the initial data because $\det A^{\Om_0}\neq 0$. The condition of $A^{\Om_0}$ being bounded away from zero is equivalent to finding some $\delta >0$ such that $\unph N^{-1}>\delta$. This holds for sufficiently small $\Om$, because by \eqref{trfalloff} we have $\unph N^{-1} = \ell^{-1} + \mathcal{O}(\Om)$, thus it suffices choosing $\delta = \ell^{-1}/2$ as a lower bound for $\unph N^{-1}$.


\subsection{Stability of Friedrich equations}\label{linstab}

Friedrich's strategy to prove that his conformal equations form a well-posed system consists in showing that they can be reduced to a symmetric hyperbolic quasilinear system of PDEs (cf. \cite{Friedrich:1986qfi}). {This in turn entails a stability result like the one we have discussed at the beginning of this appendix, which Friedrich applies to solutions close to de Sitter (see Theorem 3.3 of \cite{Friedrich:1986qfi}). We note that the paper states that same stability holds for all (weakly) asymptotically simple (i.e. admitting a conformal extension) solutions with positive $\Lambda$ (see Remark 3.4 $(ii)$ of \cite{Friedrich:1986qfi}). The initial data on the initial surface $D_{\Om_0} = \Sigma$ are required to be $H^m(\Sigma,\mathbb{R}^n)$ with $m\geq 4$, which implies regularity of the data at least $C^{m-2}(\Sigma,\mathbb{R}^n)$ and solutions $ C^{m-2}([\Om_0,\Om_1]\times \Sigma,\mathbb{R}^n)$.}

Given a solution $\unph g_0$ of Friedrich's equations, stability holds all over the unphysical manifold $\unph M$, up to and including $\scri$. 
Thus, let us consider $\unph g_0$ such that it extends from some initial Cauchy slice $D_{\Om_0}$ to $\scri^+ = \{ \Om = 0\}$. Let $u_0$ be the initial data of $\unph g_0$ at $D_{\Om_0}$. Then there is a ball $B_r(u_0)$ (as defined above in this appendix) of sufficiently small radius $r$ such that all metrics with initial data in $B_r(u_0)$ also extend to $\scri^+ = \{ \Om = 0\}$. Consider a family of data $\{u_\epsilon \} \subset B_r(u_0)$ depending smoothly on a parameter $\epsilon$ and converging to $u_0$ for $\epsilon = 0$. Then, the corresponding family of solutions $\unph g_\epsilon$ converges to $\unph g_0$ and extends to $\scri^+$. From this fact it follows that the linear fields $\delta \unph g = \frac{d \unph g_\epsilon}{d\epsilon}\mid_{\epsilon = 0}$, which are solutions of the linearized equations with background field $\unph g_0$, must also extend to $\scri^+$. 

\section{Gauss theorem}\label{appstokes}

In this appendix we make explicit the Gauss theorem in the context we are working at. Namely,
\begin{equation}\label{eqstokes}
\int_{D} |g|^{1/2} \nabla_{a} v^a = \int_{\partial D} |\gamma|^{1/2} n^{\mathrm{out}}_a\, v^a,
\end{equation}
where $n^{\mathrm{out}}_a$ is outward pointing 1-form normal to the boundary $\partial D$ of a domain $D$.

In our specific case, we consider domain $\dom_\epsilon$ defined by $\Omega>\epsilon$, where $\epsilon$ is a sufficiently small positive constant. The domain of integration is endowed with a metric $g_{ab}$. The boundary manifold $\partial \dom_\epsilon$, given by $\Om = \epsilon$ surface, has an induced spacelike metric $\gamma_{ab}$. Therefore, $n^a = N \nabla^a \Om$ is the unit timelike normal, which is outward pointing (see comment below equation \eqref{eqboundEH}) and $-n_a = -N\nabla_a\Omega$ is outward pointing normal 1-form. Then we get the Gauss theorem in the form 
\begin{equation}\label{eqstokes}
\int_{D_{\epsilon}} |g|^{1/2} \nabla_{a} v^a = -\int_{\partial \dom_\epsilon} |\gamma|^{1/2} n_a v^a.
\end{equation} 

Furthermore, recall that the {Gauss} theorem is a topological result, thus independent from the metric. Indeed, the LHS of \eqref{eqstokes} is a total derivative and can be written in terms of the unphysical metric $\unph g_{ab}$ and connection $\unph \nabla$ simply as 
\begin{equation}
\int_{\dom_\epsilon} |g|^{1/2} \nabla_{a} v^a = \int_{\dom_\epsilon} |\unph g|^{1/2} \unph \nabla_{a} v^a .
\end{equation}
This is a general identity for integrals of total derivatives, so in particular, it also holds in the boundary manifold that 
\begin{equation}\label{eqderdensity}
\int_{\partial \dom_\epsilon} |\gamma|^{1/2} D_{a} u^a = \int_{\partial \dom_\epsilon} |\unph \gamma|^{1/2} \unph D_{a} u^a ,
\end{equation}
for tangent vectors $u^a$ to $\partial \dom_\epsilon$.

\bibliography{Flux}
\end{document}